\documentclass[12pt,a4paper]{article}
\usepackage{amsmath, amssymb, amscd, amsthm, amsfonts}
\usepackage{caption}
\usepackage{subcaption}
\usepackage[final]{graphicx}
\usepackage{hyperref}
\usepackage{stackengine}
\usepackage[utf8]{inputenc}
\usepackage{natbib}
\usepackage{float}
\usepackage{xcolor}

\title{}
\author{}
\date{}

\title{Eco-evolutionary dynamics in finite network-structured populations with migration}
\begin{document}

\maketitle

\noindent
\begin{minipage}{\textwidth}\centering
$^1$Karan Pattni, $^1$Wajid Ali, $^2$Mark Broom, $^1$Kieran J Sharkey
\\$^1$Department of Mathematical Sciences, University of Liverpool
\\$^2$Department of Mathematics, City, University of London
\end{minipage}

\begin{abstract}
We consider the effect of network structure on the evolution of a population. 
Models of this kind typically consider a population of fixed size and distribution.
Here we consider eco-evolutionary dynamics where population size and distribution can change through birth, death and migration, all of which are separate processes.
This allows complex interaction and migration behaviours that are dependent on competition.
For migration, we assume that the response of individuals to competition is governed by tolerance to their group members, such that less tolerant individuals are more likely to move away due to competition.
We looked at the success of a mutant in the rare mutation limit for the complete, cycle and star networks.
Unlike models with fixed population size and distribution, the distribution of the individuals per site is explicitly modelled by considering the dynamics of the population.
This in turn determines the mutant appearance distribution for each network. 
Where a mutant appears impacts its success as it determines the competition it faces.
For low and high migration rates the complete and cycle networks have similar mutant appearance distributions resulting in similar success levels for an invading mutant.
A higher migration rate in the star network is detrimental for mutant success because migration  results in a crowded central site where a mutant is more likely to appear.
\end{abstract}

\section{Introduction}

Migration is one of the drivers of evolutionary processes.
One of the ways in which the effect of migration can be captured is  to consider a subdivided population.
Each subdivision is a unit of space that can be occupied by one or many individuals.
In ecology such models are used to study species in fragmented habitats \citep{hanski1998} such as the  fritillary butterfly \citep{wahlberg2002}.
In evolutionary game theory this enables modelling interactions between subsets of individuals \citep{broom2012}.   

Individuals can either migrate freely between these sites as in the classical island model \citep{wright1943}, or can be restricted to geographically adjacent sites as in the stepping stone model \citep{kimura1964}.
Evolutionary graph theory (EGT) \citep{lieberman2005} theoretically restricts movement using networks.
This led to interesting results where certain networks amplify the probability of a mutant type fixating in a population.
Migration in EGT occurs through replacement events where birth, death and migration are all combined such that an offspring replaces an individual in an adjacent site.
Birth, death and migration can be combined to give different replacement dynamics \citep{shakarian2012}.
Such replacement dynamics necessitate a fixed population size, which is true for models based on the EGT replacement dynamics \citep{pattni2017,yagoobi2021}.
Results therefore derived using replacement dynamics are subject to the population size being fixed.
Ecologically relevant dynamics can be obtained by considering non-replacement dynamics where birth and death are decoupled, though migration may still be coupled with either death or birth, which in effect allows for variable population size.
In \cite{pattni2021} it was shown that, in certain limiting cases, most replacement dynamics can be obtained from dynamics where death is separate but birth and migration are coupled.
This shows that models with replacement dynamics can be derived from models with variable size, and therefore the results of replacement dynamics models represent special cases of general models.
This suggests that further insights might be obtained by looking at these results in the context of non-replacement dynamics. 
As a follow up study to \cite{pattni2021}, we want to consider dynamics where birth, death and migration are all uncoupled to investigate the effect of theoretic restriction of movement using networks on the fixation probability.

In \cite{pattni2021} network structure was incorporated into the individual-based model of \cite{champagnat2006}, where the time-scale of individual level processes can be changed to consider different types of evolutionary models. 
For example, the evolution of RNA viruses \citep{grenfell2004} where evolutionary and ecological timescales overlap.
In this model, individuals reproduce asexually such that migration was coupled with birth, this is akin to dispersal in plants \citep{fournier2004} or the spread of infection \citep{rosenquist2010}.
Uncoupled migration, where individuals can freely move between sites, enables us to consider complex behaviours such as animal migration \citep{bauer2013}.
It also enables us to study social dilemmas \citep{broom2018} where assortment or grouping is required to achieve a social outcome \citep{fletcher2009}.
\cite{broom2012} presents a framework that allows complex movement behaviours to be considered in network structured populations.
The simplest case is where individuals migrate independent of their history, such that they move from one site to another with a fixed probability. 
An example of where migration is dependent on history is the Markov movement model \citep{pattni2017}.
In this case movement is a function of an individual's current group interaction and includes factors such as tolerance to group members and movement cost.
When there is high tolerance to group members, movement is essentially independent of group interactions.
On the other hand, low tolerance to group members means individuals are sensitive and more likely to move away from non-beneficial interactions.
Markov movement is a way of capturing density-dependent movement that explains a wide variety of ecological aggregations \cite{liu2016}.
In this paper we implement a version of this Markov movement where tolerance to group members plays a key role in determining whether individuals move or stay.

We start by explaining the framework in section 2, where the rare mutation limit evolutionary scenario is described.
In section 3 we provide an example of a birth-death-migration model derived from the framework where Markov movement is used.
We consider the trivial case with one site, the low migration limit and a general migration rate.
For the general migration rate we investigate the effect of migration rate and how this compares to the low migration limit.

\section{Modelling Framework}
\begin{table}[htp]
\caption{Summary: Notations for  framework, and their definitions and descriptions.} 
\centering  
\begin{tabular}{p{2cm} p{5cm} p{8cm}} %
\hline  
 Notation & Definition & Description 
\\ 
\hline
$N$
& $\geq 1$
& Number of distinct sites.
\\
$W$ 
& $W_{m,n} \geq 0$  
& Weighted $N\times N$ matrix representing network of sites.
\\
$\mathcal{U}$ 
& $\subset \mathbb{R}^{l}$  
& $l$ real-valued phenotypic traits of an individual.
\\
$\mathcal{X}$ 
& $= \{1,\dots, N\}$ 
& Set of sites an individuals can occupy.
\\
$i $ & $ = (U_{i},X_{i})$ for $U_{i} \in \mathcal{U}$ and \mbox{$X_{i} \in \mathcal{X}$}
& The traits of an individual.
\\
$I_{i}$ 
&  
& An individual with traits $i$.
\\
$\mathcal{S}$ 
& $=\{i^{m(i)}:i\in \mathcal{I},\mathcal{I}\subseteq\mathcal{U}\times\mathcal{X}\}$ 
& Multiset that gives the state of the population, where $m:\mathcal{I}\to \mathbb{Z}^+$ is the multiplicity (number of occurrences) of $i$.
\\
$\mathcal{S}_{n}$ 
& $= \{i \in \mathcal{S}:\,X_{i}=n\}$ 
& Individuals present in site $n$, therefore,  $\mathcal{S}_{n} \subseteq \mathcal{S}$.
\\
$d(i,\mathcal{S})$ 
& $\geq 0$ 
& Death rate of $I_{i}$ in state $\mathcal{S}$.\\
$b(i,\mathcal{S})$ 
&  $ \geq 0$ 
& Birth rate of $I_{i}$ in state $\mathcal{S}$.\\
$\mu(i)$ 
& $\geq 0$ 
& Probability that an offspring of $I_{i}$ carries a mutation.
\\
$M(u, v)$ 
& $\geq0$
& Probability that offspring has mutated trait $v$ when parent has trait $u$.
\\
$m(i,x,\mathcal{S})$
& $\geq0$ 
& Migration rate of $I_i$ to site $x$ in state $\mathcal{S}$. 
\\
$\phi$ 
& $:\mathcal{S}\rightarrow \mathbb{R}$
& Real-valued bounded function that acts on the state of the system.\\
$\mathcal{L}$ 
&  
& Markov process generator, it describes how the expected value of $\phi$ changes for an infinitesimal time interval.
\\
$h_\mathcal{A}(\mathcal{S})$
& $\in[0,1]$
& Probability of starting in state $\mathcal{S}$ and hitting state in set $\mathcal{A}$. 
\\
\hline  
\end{tabular} 
\label{tab:Champagnat}
\end{table}

We assume that the individuals in the population are spread in distinct but connected sites of a fixed network. The sites of the network can have no, one or many individuals at a given time.  The population size and composition can change due to birth and death, whereas migration changes the distribution of individuals across the network of sites.

To mathematically describe such populations, we use the \cite{champagnat2006} model with network structure as described in \cite{pattni2021}.
Individuals can have $l$ real-valued traits contained within the set $\mathcal{U}\subset \mathbb{R}^l$.
The sites that individuals can occupy is given by set $\mathcal{X}=\{1,\ldots,N\}$.
The characteristics of an individual are given by $i=(U_i,X_i)$, where $U_i\in \mathcal{U}$ and $X_i\in\mathcal{X}$.
An individual with characteristics $i$ is denoted by $I_i$. 
The state of the population is given by a multi-set $\mathcal{S}$, which means that for each individual with characteristics $i$ there is a copy of $i$ in $\mathcal{S}$.
Formally we write this as $\{i^{m(i)}:i\in \mathcal{I},\mathcal{I}\subseteq\mathcal{U}\times\mathcal{X}\}$ where $m:\mathcal{I}\to \mathbb{Z}^+$ is the multiplicity (number of occurrences) of $i$.
Individuals in the same site are given by set $\mathcal{S}_n=\{i\in\mathcal{S}:X_i=n\}$. 

The connections between sites are given by a directed and weighted network represented by a matrix $W$ with entries $W_{m,n}\ge 0$. 
An individual can move from site $m$ to $n$ if site $m$ is connected to site $n$; that is, $W_{m,n}>0$.
In \cite{pattni2021}, birth and movement were coupled such that an offspring can be placed onto a connected site.
Here we consider uncoupled birth and movement.
Individuals are assumed to reproduce asexually such that they place their offspring on the same site.
The rate at which individual $I_i$ gives birth is given by $b(i,\mathcal{S},W)$.
If there is no mutation, the offspring of individual $I_i$ has characteristics $i=(U_i,X_i)$. 
With probability $\mu(i)$, individual $I_i$ gives birth to an offspring with mutation.
In this case, the probability that the $I_i$ gives birth to an offspring with trait $u$ is $M(U_i,u)$ such that all mutations are contained within $\mathcal{U}$, that is, $M(U_i,u)=0$ if $u\notin \mathcal{U}$. 
The rate at which individual $I_i$ dies is given by $d(i,\mathcal{S},W)$. 
The rate at which individual $I_i$ moves to site $x$ is given by $m(i,x,\mathcal{S},W)$. 
Since the network structure $W$ is assumed to be fixed, we use $b(i,\mathcal{S})$, $d(i,\mathcal{S})$ and $m(i,x,\mathcal{S})$ for the birth, death and movement rate respectively.

The evolution of the population is described by a continuous time Markov process. 
The generator $\mathcal{L}$ that acts on real bounded functions $\phi(\mathcal{S})$ that describes the infinitesimal dynamics of the state of the population at time $t$ is given by 
\begin{equation}\label{equ:Simple_Movement}
\begin{split}
\mathcal{L}\phi(\mathcal{S}) = 
    &\sum_{i\in \mathcal{S}} 
        [1 - \mu(i)]
        b(i,\mathcal{S})
        [\phi(\mathcal{S} \cup \{i\}) - \phi(\mathcal{S})] \\
    & + \sum_{i \in \mathcal{S}} 
        \mu(i)b(i,\mathcal{S}) 
        \int_{\mathbb{R}^{l}}
        [\phi(\mathcal{S}\cup \{(u,X_i)\}) - \phi(\mathcal{S})]M(U_i,u)du\\
    & + \sum_{i\in \mathcal{S}} d(i,\mathcal{S})[\phi(\mathcal{S}\backslash \{i\}) - \phi(\mathcal{S})]\\
    & +\sum_{i \in \mathcal{S}}\sum_{x\in \mathcal{X}} m(i,x,\mathcal{S})[\phi(\mathcal{S} \cup \{(U_i,x)\}\backslash\{i\} - \phi(\mathcal{S})].
\end{split}    
\end{equation}
The first line describes birth without mutation, the second line describes birth with mutation, the third line describes death and the fourth line describes migration.

For the Markov process described by the infinitesimal dynamics in equation \eqref{equ:Simple_Movement}, the key quantity we are interested in is the hitting probability. 
The probability $h_\mathcal{A}(\mathcal{S})$ of starting in state $\mathcal{S}$ and hitting a state in set $\mathcal{A}$ is calculated as follows 
\begin{align}
     \mathcal{L}h_\mathcal{A}(\mathcal{S}) = 0 
\end{align}
with boundary condition $h_\mathcal{A}(\mathcal{S}) = 1$ for $\mathcal{S} \in \mathcal{A}$ (see \cite{pattni2021}, Appendix A).

\subsection{Evolution in the Rare Mutation Limit}
In the rare mutation limit we assume that $\mu(i)=\mu\to 0\ \forall i$ so that the population evolves through adaptive sweeps \citep{gerrish1998}. 
This means that, prior to a mutation arising, the population is homogeneous with all individuals having the same traits.
This is because, when a mutation appears, either all individuals with the mutation (referred to as mutants and denoted $M$) die out or all individuals without the mutation (referred to as residents and denoted $R$) die out prior to another mutation arising.
There can therefore be at most two types in the population, a type $R$ and a type $M$. 
Let $\mathcal{U}=\{R,M\}$, then the set of states where all individuals are residents is given by $\mathcal{R}=\{\mathcal{S}:U_i=R,\ \forall i\in \mathcal{S}\}$; similarly the set of all states with mutants is given by $\mathcal{M}=\{\mathcal{S}:U_i=M,\ \forall i\in \mathcal{S}\}$.
The dynamics of the system can therefore be described without the mutation step; that is, equation \eqref{equ:Simple_Movement} simplifies to
\begin{equation}
\label{eq:simplified generator}
\begin{split}
\mathcal{L}\phi(\mathcal{S}) = 
    &\sum_{i\in \mathcal{S}} 
        b(i,\mathcal{S})
        [\phi(\mathcal{S} \cup \{i\}) - \phi(\mathcal{S})] \\
    & + \sum_{i\in \mathcal{S}} d(i,\mathcal{S})[\phi(\mathcal{S}\backslash \{i\}) - \phi(\mathcal{S})]\\
    & +\sum_{i \in \mathcal{S}}\sum_{x\in \mathcal{X}} m(i,x,\mathcal{S})[\phi(\mathcal{S} \cup \{(U_i,x)\}\backslash\{i\} - \phi(\mathcal{S})].
\end{split}    
\end{equation}

When the population is in a homogeneous state with all residents prior to a mutant arising, i.e.~$\mathcal{S}\in\mathcal{R}$, we are interested in determining the state in which a mutant appears.
Let $\pi(\mathcal{S})$ be the probability that the population is in state $\mathcal{S}$. 
This can be calculated using equation \eqref{eq:simplified generator} as follows
\begin{align}
    \mathcal{L}\pi(\mathcal{S}) = 0, \quad \mathcal{S}\in\mathcal{R}
\end{align}
with normalising condition
\begin{align}
    1=\sum_{\mathcal{S}\in\mathcal{R}}\pi(\mathcal{S}).
\end{align}
The probability $p_{x,\mathcal{S}}$ that a mutant appears in site $x$ in state $\mathcal{S}$ is proportional to the number of individuals in site $x$; that is,
\begin{align}
    p_{x,\mathcal{S}} = \frac{|\mathcal{S}_x|}{|\mathcal{S}|}\pi(\mathcal{S}). 
\end{align}
Note that whether a unique solution to $\pi(\mathcal{S})$ exists depends upon the definition of birth, death and movement.

Once a mutation arises, the type that remains is said to have fixated in the population, and we are interested in the probability of mutants fixating.
This is calculated by solving the hitting probability using equation \eqref{eq:simplified generator} as follows 
\begin{align}
    \mathcal{L}h_\mathcal{M}(\mathcal{S}) = 0
    \label{eq_hitting_prob}
\end{align}
with boundary conditions $h_\mathcal{M}(\mathcal{S}) = 1$ for $\mathcal{S} \in \mathcal{M}$ and $h_\mathcal{M}(\mathcal{S}) = 0$ for $\mathcal{S} \in \mathcal{R}$.
To be precise with terminology, we refer to the fixation probability as the probability of one initial mutant fixating.
Since there are multiple states with one mutant, we calculate the average fixation probability as follows
\begin{align}
    \rho = 
    \sum_{\mathcal{S} \in \mathcal{R}}
    \sum_{x\in\mathcal{X}}
    p_{x,\mathcal{S}}
    h_\mathcal{M}(\mathcal{S}\cup\{(M,x)\})  
\end{align}
where $p_{x,\mathcal{S}}$ is the mutant appearance distribution, i.e.~the probability that a mutant appears in site $x$ when the population is in state $\mathcal{S}$.

\section{Birth-Death-Migration Model}
To apply the modelling framework, we consider a birth-death-migration model that we can use to calculate the fixation probability.
The birth rate is considered to be fixed and depends only on the type of individual:
\begin{align}
    b(i,\mathcal{S}) = \beta_{U_i}. 
\end{align}
The death rate is given by  
\begin{align}
    d(i,\mathcal{S}) = \delta_{U_i}+\sum_{j\in\mathcal{S}_{X_i}\setminus \{i\}}\gamma_{U_i,U_j}
\end{align}
where $\delta_u$ is the natural death rate of a type $u$ individual and $\gamma_{u,v}$ is the death rate of a type $u$ individual when competing with a type $v$ individual.

We assume that individuals move with migration rate $\lambda>0$.
Where they move to will depend upon the structure of the network given by $W$.
We assume that $W_{x,x}=0$ and $\sum_{y\in \mathcal{X}} W_{x,y} = 1 \ \forall x\in\mathcal{X}$; that is, all diagonal elements of $W$ are zero and $W$ is right-stochastic. 
This means that $W_{x,y}$ is the probability of migrating from site $x$ to $y$.
In \cite{pattni2018} a density-dependent movement function was considered where the movement of individuals is determined by their staying propensity and tolerance to other group members.
The staying propensity can capture the cost of movement such that if the cost of movement is high, individuals are likely to have a higher staying propensity.
Tolerance controls how sensitive individuals are to their group members, such that the higher the tolerance the more likely they are to stay.
These aspects were captured using a sigmoid function in \cite{pattni2018}, we consider an adapted version of this as follows
\begin{align}
    m(i,x,\mathcal{S})=
    \left(
    1-
    \frac{\alpha}
    {\alpha + (1-\alpha)\tau^{g(i,\mathcal{S}_{X_i})}}
    \right)
    \lambda
    W_{X_i,x}.
\end{align}
The staying propensity is controlled by $\alpha\in[0,1]$, which is the probability that individual $i$ stays in its current site.
If $\alpha=1$ then individuals will stay regardless of the interactions they have with other members of the population.
The benefit of individual $i$ being in group $\mathcal{S}_{X_i}$ is given by $g(i,\mathcal{S}_{X_i})$; if this is positive then being present in the group is beneficial.
The effect of group benefit is determined by tolerance to other group members, which is given by $\tau\in(0,1)$. 
The following two limiting cases for tolerance are of note.
\begin{enumerate}
    \item \textbf{High tolerance to group members ($\tau\to 1$):} In this case individuals move independently of one another as the benefit of being in a group has virtually no impact. 
    The movement rate for high group tolerance (HGT) is given by
\begin{align}
    m^\text{HGT}(i,x,\mathcal{S}) =(1-\alpha)\lambda W_{X_i,x}. 
    \label{eq_hgt_general}
\end{align}
    \item \textbf{Low tolerance to group members ($\tau\to 0$):}
In this case the focal individual will stay if its current group is beneficial, i.e.~if $g(i,\mathcal{S}_{X_i})>0$, but can move otherwise. 
The movement rate for low group tolerance (LGT) is given by
\begin{align}
    m^\text{LGT}(i,x,\mathcal{S}) =
        \begin{cases}
            \lambda W_{X_i,x} & g(i,\mathcal{S}_{X_i}) < 0, \\
            (1-\alpha)\lambda W_{X_i,x} & g(i,\mathcal{S}_{X_i})=0,\\
            0 & g(i,\mathcal{S}_{X_i}) > 0.
        \end{cases}
    \label{eq_lgt_general}
\end{align}
\end{enumerate}

\subsection{Example of birth-death-migration model}
As an initial application of the birth-death-migration model, we consider a simple example that enables us to obtain analytical results in certain limiting cases. The simplifications used are described as follows.

Different types of individuals differ in terms of their birth rate only and cannot die naturally.
We set the birth rate of a resident to $\beta_R=1$ and mutant to $\beta_M=2$, unless specified otherwise. 
No natural death means that $\delta_u=0$ for $u\in\{M,R\}$.
This means that extinction events are avoided.
An alternative way of dealing with extinction events is to reseed the population, however, we avoid this technicality for now.
Individuals therefore die due to competition with an identical rate for all paired types, i.e.~$\gamma_{u,v}=\gamma,\ \forall u,v$.

For density-dependent movement, we only consider the limiting cases of low and high group tolerance, and assume that the probability of staying is small ($\alpha\ll 1$) so that its impact is negligible. 
For the group benefit function it is assumed that the focal individual benefits when they are alone; that is,
\begin{align}
    g(i,\mathcal{S}_{X_i}) = 
    \begin{cases}
        \phantom{-}1 & |\mathcal{S}_{X_i}| = 1, \\
        -1 &  |\mathcal{S}_{X_i}| > 1.
    \end{cases}
\end{align}
This binary setting where the focal individual prefers being alone to any other group, suppresses nuanced group effects where, for example, group preference changes gradually with group size. 
However, it still impacts the migration of individuals which is our main focus.
For low group tolerance (LGT), equation \eqref{eq_lgt_general} simplifies to 
\begin{align}
    m^\text{LGT}(i,x,\mathcal{S}) =
        \begin{cases}
            \lambda W_{X_i,x} & |\mathcal{S}_{X_i}| > 1, \\
            0 & |\mathcal{S}_{X_i}| = 1
        \end{cases}
        \label{eq_lgt}
\end{align}
which specifies that individuals will not migrate when they are alone.
For high group tolerance (HGT), equation \eqref{eq_hgt_general} simplifies to 
\begin{align}
    m^\text{HGT}(i,x,\mathcal{S}) =\lambda W_{X_i,x}
    \label{eq_hgt}
\end{align}
which specifies that individuals will migrate regardless of being in a group or not.

\begin{figure}
     \centering
     \includegraphics{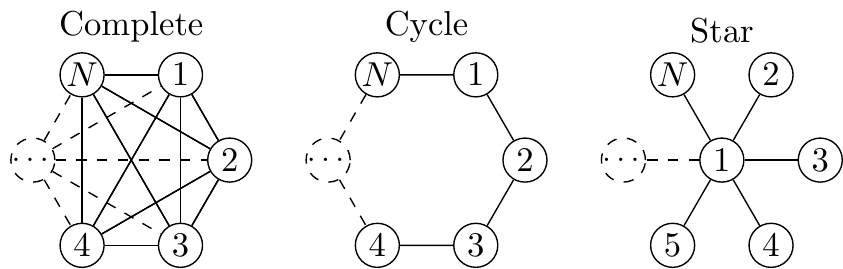}
     \caption{Networks considered in this paper. Each node represents a site. $N$ is the total number of sites. Each edge represents an incoming and outgoing weighted edge whose weights are given by $W$. Edges represent the permitted migration routes of individuals. In the star network site 1 is called the centre and sites 2 to N are called leaf sites.}
     \label{fig:networks}
 \end{figure}

The complete ($W^\bullet$), cycle ($W^\circ$) and star ($W^\star$) networks will be considered, they are illustrated in figure \ref{fig:networks}.
For each network, $W_{ii}^\bullet=W_{ii}^\circ=W_{ii}^\star=0\ \forall i \in\mathcal{X}$ and the non-zero weights are as follows
\begin{align}
    \left.
    \begin{array}{l}
    \text{Complete: } W_{ij}^\bullet=1/(N-1),\ i\ne j \text{ and } i,j\in\mathcal{X},\\
    \text{Cycle: } W_{i,i+1}^\circ=W_{j,j-1}^\circ=W_{1,N}^\circ=W_{N,1}^\circ=\frac{1}{2},\ i=1,\ldots,N-1 \text{ and } j =2,\ldots,N,\\
    \text{Star: }
    W_{1,i}^\star=\frac{1}{N-1},\ W_{i,1}^\star=1,\ i={2,\ldots,N-1}.
    \end{array}
    \right\}
    \label{eq:weights}
\end{align} 
Due to the properties of the networks chosen, they will provide us with a base understanding of the birth-death-migration model without the need to run lengthy simulations across a wide range of networks.
The complete network is the benchmark case.
The complete and cycle networks are circulations \citep{lieberman2005}, which means that the sum of the incoming and outgoing weights for each site are the same, i.e.~$\sum_{i} W_{i,j}=\sum_{j}W_{j,i} \quad \forall i,j\in\mathcal{X}$.
This property is used to derive the Circulation theorem \citep{lieberman2005}, which states circulation networks have the same fixation probability.
We may therefore be able to use this property to extend our results to other circulation networks.
For the weights we have chosen, the star network can amplify selection \citep{lieberman2005,pattni2021}, which means that the fixation probability of a mutant is greater than in the complete network, so we can check whether this is still the case when there is uncoupled movement.

Details regarding the simulation of the birth-death-migration model example are given in the appendix.
The simulations were carried out using the HTCondor distributed computing system \citep{thain2005}.
 
\subsection{Special case with single site}
We first consider the case where there is one site.
This will allow us to understand the intra-site dynamics.
For the birth-death-migration model we can analytically calculate the stationary distribution of a homogeneous population (i.e.~all residents or all mutants).
Let $\pi_n^u=\mathbb{P}(\mathcal{S}=\{(u,1)^n\})$ be the probability that there are $n$ individuals of type $u$ in the population.
Recall that the population cannot go extinct because we have assumed that the natural death rate is zero in this example (death only occurs by competition). 
The homogeneous population is therefore described by a reversible Markov process and we can obtain $\pi_n^u$ using the detailed balance equations, which states that the transition rates do not change when time is reversed. 
In particular, the rate at which we transition from state $n$ to $n-1$ due to a death event is the same as transitioning from state $n-1$ to $n$ due to a birth event.
In a state with $n$ individuals each individual dies with rate $\gamma(n-1)$ and in a state with $n-1$ individuals each individual gives birth with rate $\beta$, so for $n\ge2$ the detailed balance equations give
\begin{align}
	n\gamma (n-1)\pi_n^u 
	& = (n-1)\beta_u \pi_{n-1}^u\nonumber
 \end{align}
 which simplifies to 
 \begin{align}
	\pi_{n}^u
	&= \frac{1}{n} \frac{\beta_u}{\gamma} \pi_{n-1}^u
    \nonumber
\end{align}
and through recursion we obtain
\begin{align}
	\pi_{n}^u &= \frac{(\beta_u/\gamma)^{n-1}}{n!}  \pi_{1}^u
	.
 \label{eq:recursion}
\end{align}
Using the fact that the stationary probabilities sum to 1 (i.e.~$1=\sum_{n=1}^\infty \pi_n$) and, setting $x=\beta_u/\gamma$ for brevity,
we have that
\begin{align*}
	1&=
	\sum_{n=1}^\infty \frac{x^{n-1}}{n!}  \pi^u_{1}
	= \frac{  \pi^u_{1}}{x}\sum_{n=1}^\infty \frac{x^{n}}{n!}
	= \frac{  \pi^u_{1}}{x}\left(-1+\sum_{n=0}^\infty \frac{x^{n}}{n!}\right)
	= \frac{  \pi^u_{1}}{x}\left(-1+e^x\right),
\end{align*}
which gives
\begin{align*}
 \pi^u_1 &= \frac{(\beta_u/\gamma)}{e^{\beta_u/\gamma}-1}.
\end{align*}
Substituting $\pi_{1}^u$ into equation \eqref{eq:recursion} gives us the stationary probability,
\begin{align*}
	\pi_{n}^u &= \frac{(\beta_u/\gamma)^n}{n!}\frac{1}{e^{\beta_u/\gamma}-1}. 
\end{align*}
Using the stationary probability we calculate the expected type $u$ population size as follows
\begin{align*}
	\sum_{n=1}^\infty n \pi_{n}^u
	= \frac{1}{e^{\beta_u/\gamma}-1}\sum_{n=1}^\infty n\frac{(\beta_u/\gamma)^n}{n!}
	= \frac{(\beta_u/\gamma)e^{\beta_u/\gamma}}{e^{\beta_u/\gamma}-1}
	= \frac{\beta_u/\gamma}{1-e^{-\beta_u/\gamma}}.
\end{align*}
The appearance of a mutant is proportional to the number of resident individuals in a given state. The probability that an initial mutant appears in a state with $n$ residents is therefore given by 
\begin{align*}
\mu_n^\text{Init} = 
\frac{n\pi_n^u}{\sum_{i=1}^\infty i\pi_i^u} = 
n\frac{(\beta_R/\gamma)^n}{n!}\frac{1}{e^{\beta_R/\gamma}-1}\left/\frac{(\beta_R/\gamma)e^{\beta_R/\gamma}}{e^{\beta_R/\gamma}-1}\right.=
\frac{(\beta_R/\gamma)^{n-1}}{(n-1)!e^{\beta_R/\gamma}}.
\end{align*}
Using this probability, the average fixation probability of a mutant in a single site is then given by
\begin{align}
	{\rho}^\text{Single} = \sum_{n=1}^\infty \mu_n^\text{Init}h_{\mathcal{M}}(\{(R,1)^n,(M,1)\}).
	\label{eq_fixation_single}
\end{align}
In this case, the hitting probability can be calculated by solving equation \eqref{eq_hitting_prob} when we limit the birth rate as follows 
\begin{align}
    b(i,\mathcal{S}) = \begin{cases}
		\beta_{U_i} & |\mathcal{S}| < K,\\
		0 & |\mathcal{S}| \ge K
	\end{cases}
    \label{eq_limit_birthrate}
\end{align}  
where $K$ is chosen to be large enough so that $\mathbb{P}(|\mathcal{S}|\ge K)= 0$.
This means that the maximum population size is $K$ and the total number of states is $(K+1)^2$, because these are the total number of combinations of mutants and residents that sum to $\le K$.  

\begin{figure}[ht]
    \centering
    \includegraphics[width=0.45\linewidth]{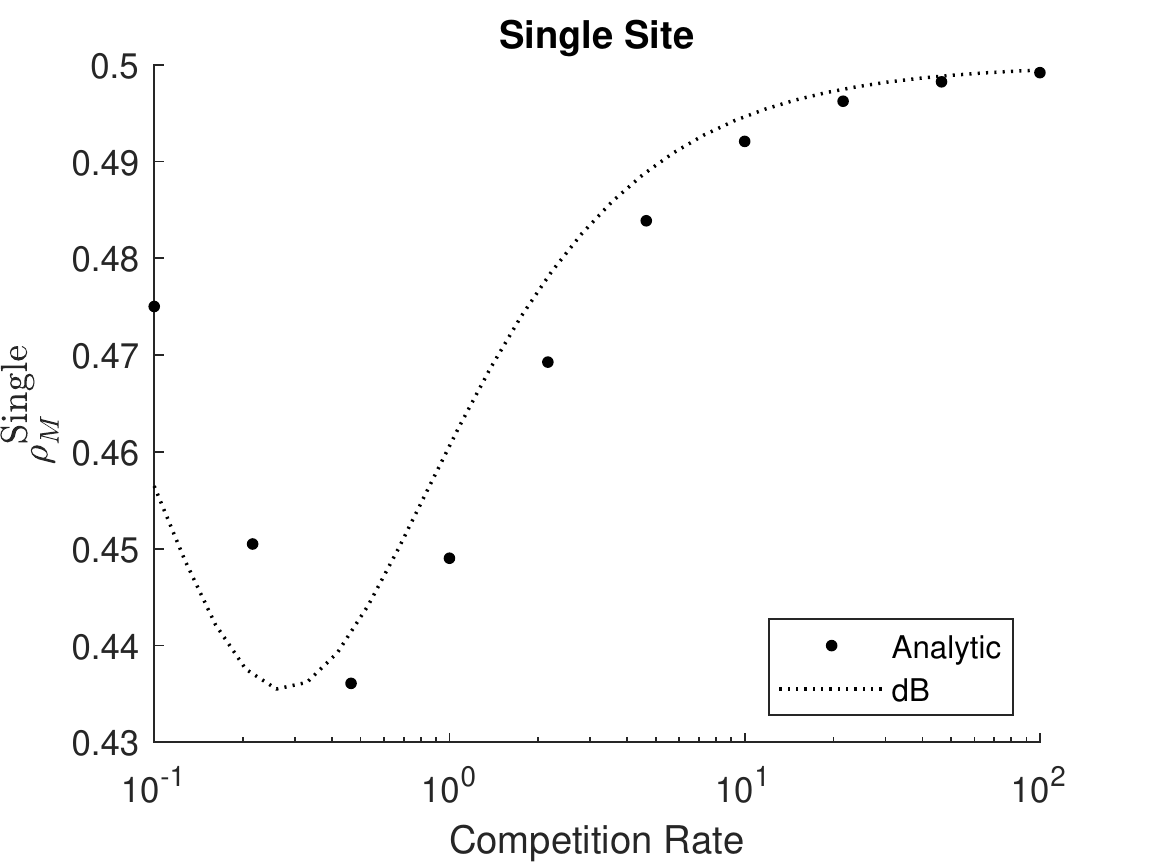}
    \caption{\label{fig_single_site}
        Fixation probability of a mutant in a single site. 
        Exact fixation probability is given by `Analytic' plot, which is calculated using equation \eqref{eq_fixation_single}.
        Approximation using dB EGT dynamics is given by `dB' plot, calculated using equation \eqref{eq_fixation_dB_approx}.
    }
 \end{figure}

Figure \ref{fig_single_site} shows the effect of competition rate on the fixation probability of a mutant in a single site  ($\rho^\text{Single}$), which is calculated by solving for $h$ using equation \eqref{eq_hitting_prob}. 
To understand the behaviour here, we look at a comparable model with fixed population size. 
We find that it resembles death-Birth (dB) EGT dynamics, where an individual is randomly chosen for death and is then replaced by an offspring of an individual who is selected for birth proportional to their fitness (hence the uppercase in dB indicates selection).
The fixation probability for dB EGT dynamics \citep{kaveh2015,hindersin2015} is given by
\begin{align}
    \rho^\text{dB}(N,r)=\frac{N-1}{N}\frac{1-\frac{1}{r}}{1-\frac{1}{r^{N-1}}}
     = \frac{N-1}{N}\rho^\text{Moran}(N-1,r)
\end{align}
where $N$ is the number of individuals, $r$ is the relative fitness of a mutant to a resident and $\rho^\text{Moran}$ is the Moran probability \citep{moran1959}.
By specifying a value for $N$ and $r$, we can use $\rho^\text{dB}$ to approximate $\rho^\text{Single}$.  
In \cite{pattni2021} it was shown that fitness in dB EGT dynamics is proportional to the birth rate of individuals, we therefore set the relative fitness to  $r=\beta_M/\beta_R$.
To set $N$, we assume that with probability $\mu_n^\text{Init}$ a  mutant arises in a population with $n$ residents, so $N=n+1$.
The maximum resident population size is set to $K$ such that having a population size $\ge K$ tends to 0. 
Putting this together gives,
\begin{align}
    \rho^\text{Single}_M \approx \sum_{n=1}^K
    \mu_n^\text{Init}
    \rho^\text{dB}(n+1,{\beta_M}/{\beta_R}).
    \label{eq_fixation_dB_approx}
\end{align}
In figure \ref{fig_single_site} we see that, even though there is some discrepancy, similar behaviour is observed with both dB dynamics and the birth-death-migration model. 
The discrepancy between the two is due to fluctuating population size in the birth-death-migration model, which allows the population to be updated via a birth or death.
With dB dynamics, the population size is fixed and can only be updated via a death-Birth event.
Note that the discrepancy increases as the population size increases (competition rate decreases).
Further insight can be obtained by looking at the components of $\rho^\text{dB}$.
The component $(N-1)/N$ in $\rho^\text{dB}$ is the probability that the initial mutant is not chosen to randomly die.
This component dominates when the population size is small since the chance of the initial mutant randomly dying is higher.
The component $\rho^\text{Moran}(N-1,r)$ captures the probability that the initial mutant fixates provided it does not randomly die.
This component dominates as the population size gets larger since the probability of the initial mutant randomly dying decreases.
This captures the behaviour observed for the single site case as follows.
As the competition rate increases, the population size decreases and, therefore, survival dictates the fixation probability of a mutant.
For a high competition rate, $\rho^\text{Single}$ converges to $\frac{1}{2}$ as the expected resident population size converges to $1$.
As the competition rate decreases, which increases the population size, the ability to reproduce (or fecundity) dictates the fixation probability of a mutant.
The dip and recovery we see in $\rho^\text{Single}$ observed in figure in \ref{fig_single_site} as we switch from fecundity dominating to survival dominating with a decreasing competition rate is due to the birth rate of mutants we have chosen ($\beta_M=2$). 
Changing the birth rate can alter this switching behaviour.

\subsection{The Low Migration Limit}
We now return to the multiple site case and consider the low migration rate limit ($\lambda \to 0$).
In this case, an initial mutant that appears on a site will die out or fixate before a migration event happens and therefore each site can be viewed as either a resident or mutant site prior to another migration event.
This approach has been previously used in the case of fixed population size within sites \citep{pattni2017,yagoobi2021}.
The probability that a mutant fixates is then a two-step process. 
First, an initial mutant appears and fixates in a single site. 
Second, mutants then spread until they fixate in the population.
The probability in the first step is given by $\rho^\text{Single}$  for both low and high group tolerance.
For low group tolerance we can obtain an analytic expression for the probability in the second step, but it is difficult for high group tolerance as we can have empty sites.
We proceed by deriving an analytic expression for low group tolerance. 
 
The rate $J^u_{x,y}$ at which a type $u$ individual migrates from site $x$ to $y$, is proportional to the expected number of individuals in site $x$ who can migrate multiplied by the migration rate.
In the case of low group tolerance (equation \eqref{eq_lgt}), the rate $J^u_{x,y}$ is given by
\begin{align}
    J_{x,y}^u=\lambda W_{x,y}\sum_{n=2}^\infty n\pi_{n,x}^u
    =\lambda W_{x,y}
    \left(
        \frac{\beta_u/\gamma}{1-e^{-\beta_u/\gamma}}
        -\frac{\beta_u/\gamma}{e^{\beta_u/\gamma}-1}
    \right)
    =\lambda W_{x,y}\beta_u/\gamma.
\end{align}
Note that $\pi^u_{n,x}=\pi^u_{n}$ where the subscript for the site is included for clarity.
To calculate the fixation probability of a type $u$ immigrant arriving in site $x$, we need to account for the number of individuals currently present in site $x$. 
With probability $\pi_{n,x}^v$ there are $n$ type $v$ individuals present in site $x$ and, therefore, the average fixation probability of a type $u$ immigrant in site $x$ is 
\begin{align}
    \rho_{u,x}^\text{Mig}&=\sum_{n=1}^\infty \pi_{n,x}^vh_{\mathcal{F}(u)}(\{(v,x)^n,(u,x)\})\nonumber\\
    & = \sum_{n=1}^\infty
    \frac{(\beta_v/\gamma)^n}{n!}\frac{1}{e^{\beta_v/\gamma}-1}
    h_{\mathcal{F}(u)}(\{(v,x)^n,(u,x)\})
\end{align}
where $v\in\{M,R\}\setminus \{u\}$ and $\mathcal{F}$ gives the state that we fixate in, that is, $\mathcal{F}(M)=\mathcal{M}$ for mutant fixation, and $\mathcal{F}(R)=\mathcal{R}$ for resident fixation.
Similar to obtaining the solution of $\rho^\text{Single}$, 
in $\rho^\text{Mig}$ we can solve $h$ using equation \eqref{eq_hitting_prob} by limiting the birth rate (equation \eqref{eq_limit_birthrate}).
Let $s\subseteq\{1,\ldots,N\}=\mathcal{X}$ represent a state of the population such that site $x$ where $x\in s$ is a site occupied by mutants and a site $y$ where $y\notin s$ represents a resident site.
We can now define the probability that mutants fixate at the site level for low group tolerance and in the low migration limit as follows
\begin{align}
\rho_{s}^\text{Low Mig} = \sum_{s'\subset\mathcal{X}}\frac{Q_{ss'}}{q_s}\rho_{s'}^\text{Low Mig}
\label{eq_fixation_low_mig}
\end{align}
with boundary conditions $\rho^\text{Low Mig}_{\emptyset}=0$ and $\rho^\text{Low Mig}_\mathcal{X}=1$, where $Q_{ss'}$ is the transition rate from state $s$ to $s'$, which is given by
\begin{align*}
    Q_{ss'}=
    \begin{cases}
        \displaystyle
        \sum_{x\notin s}
        J^R_{x,y}\rho_{R,y}^\text{Mig} & 
        \text{if }s'=s\setminus\{y\} \text{ for } y \in s,\\
        \displaystyle
        \sum_{x\in s}
        J^M_{x,y}\rho_{M,y}^\text{Mig} & 
        \text{if }s'=s\cup\{y\} \text{ for } y \notin s,\\
        0 &\text{otherwise,}
    \end{cases}
\end{align*}
and $q_s$ is the rate of transitioning away from state $s$, that is
\begin{align}
    q_s = \sum_{s' \subseteq \mathcal{X}} Q_{ss'}.
    \nonumber
\end{align}
The average fixation probability of a mutant for low group tolerance and in the low migration limit is then given by
\begin{align}
    \rho^\text{LGT} = \sum_{x\in\mathcal{X}}p_x\rho_{M,x}^\text{Single}\rho^\text{Low Mig}_{\{x\}}
    \label{eq_fixation_low_mig_lgt}
\end{align}
where $\rho^\text{Single}_{M,x}=\rho^\text{Single}_{M}$, but have included site index for clarity, and $p_x$ is the probability a mutant appears in site $x$, which is proportional to the expected number of individuals in a site, that is,
\begin{align}
    p_x = \frac{\sum_{n=1}^\infty n\pi^R_{n,x}}{
        \sum_{y\in\mathcal{X}}\sum_{n=1}^\infty n\pi_{n,y}^R
    }=\frac{1}{N}.
    \nonumber
\end{align}
Note that the intra-site dynamics are homogeneous, i.e.~$\pi^R_{n,x}=\pi_{n,y}^R\ \forall x,y$, so the mutant appearance distribution is uniform.
Since we have homogeneous intra-site dynamics, we can use the circulation theorem \citep{lieberman2005} to calculate $\rho^\text{Low Mig}$ in circulation networks. 
The theorem is derived from the property that circulation networks have a constant forward bias for all transitory states (both residents and mutants exist) \citep{lieberman2005,pattni2015a}, and it therefore states that the fixation probability can be obtained by the Moran probability
\begin{align}
	\rho^\text{Moran}(N,f)=\frac{1-\frac{1}{f}}{1-\frac{1}{f^N}},
	\label{eq_moran}
\end{align}
where $N$ is the number of sites and $f$ is the forward bias.
For a transitory state $s$ the forward bias $f$ is given by the rate of mutants increasing divided by the rate of residents decreasing; that is, 
\begin{align}
    f = 
    \frac{
        \sum_{x\in s}
        J^M_{x,y}\rho_{M,y}^\text{Mig}
    }{ 
        \sum_{x\notin s}
        J^R_{x,y}\rho_{R,y}^\text{Mig}
    }.
	\label{eq:forward_bias}
\end{align}
Therefore, in the case of circulation networks, we can substitute $\rho_{\{x\}}^\text{Low Mig}$ with $\rho^\text{Moran}(N,f)$ for all $x$, so equation \eqref{eq_fixation_low_mig_lgt} simplifies to  
\begin{align}
	\rho^\text{LGT Circ} = {\rho}^\text{Single}_M\rho^\text{Moran}(N,f).
	\label{eq_analytic}
\end{align}
For the star network, $\rho^\text{LGT}$ (equation \eqref{eq_fixation_low_mig_lgt}) can be calculated using the formula in \cite{broom2008}. 

\begin{figure}
\centering
\includegraphics[width=0.45\linewidth]{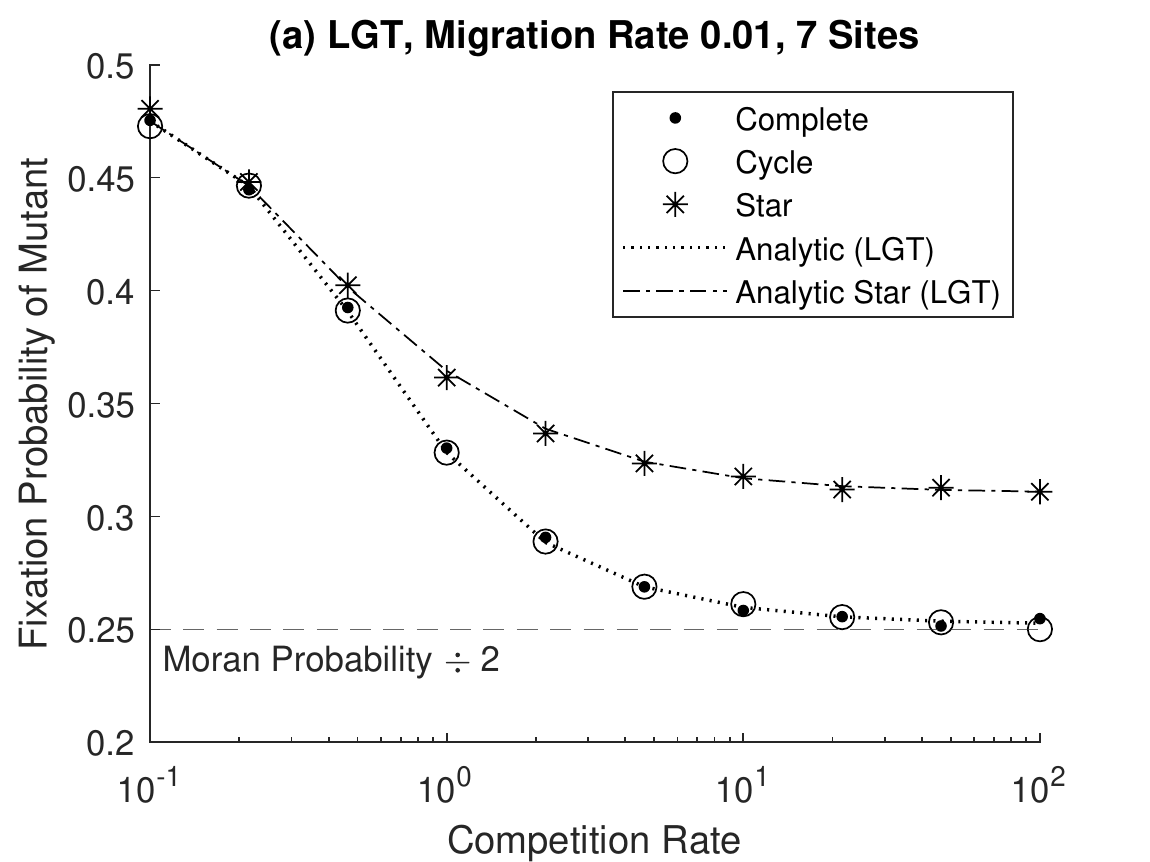}
\includegraphics[width=0.45\linewidth]{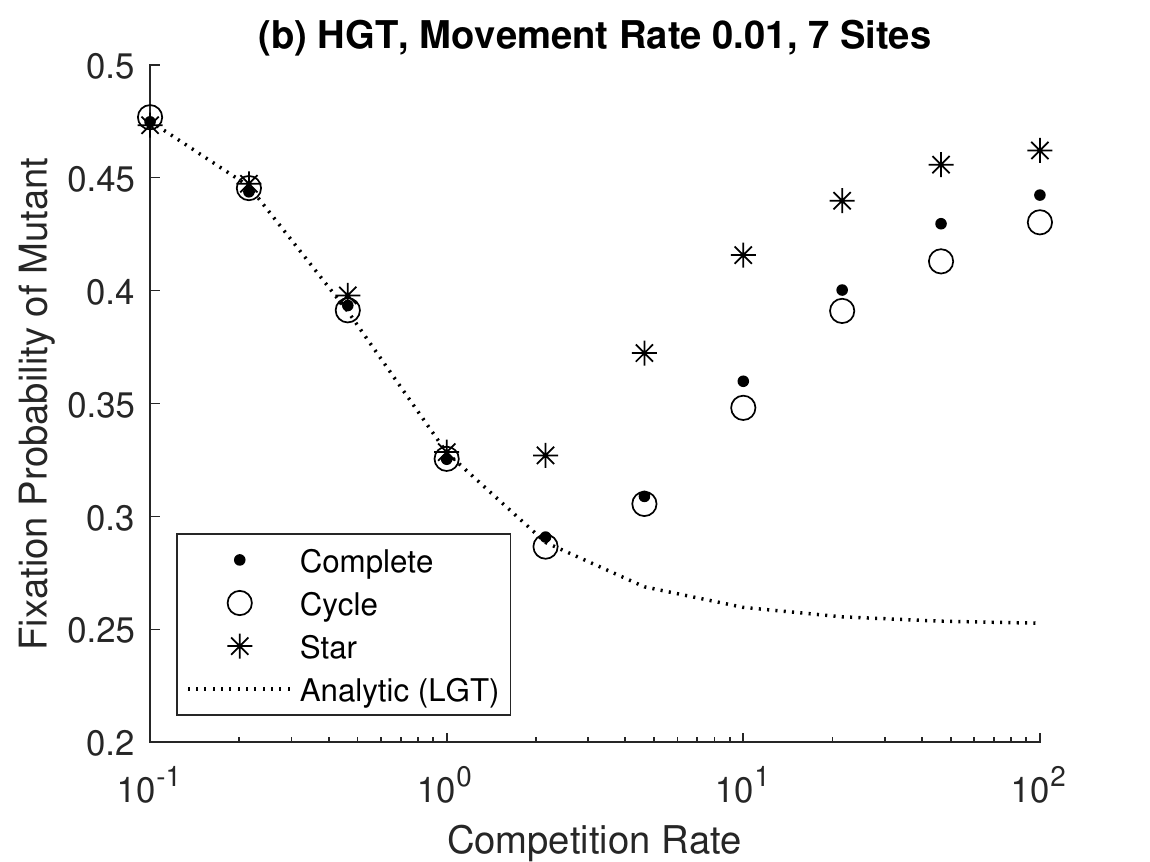}
\includegraphics[width=0.45\linewidth]{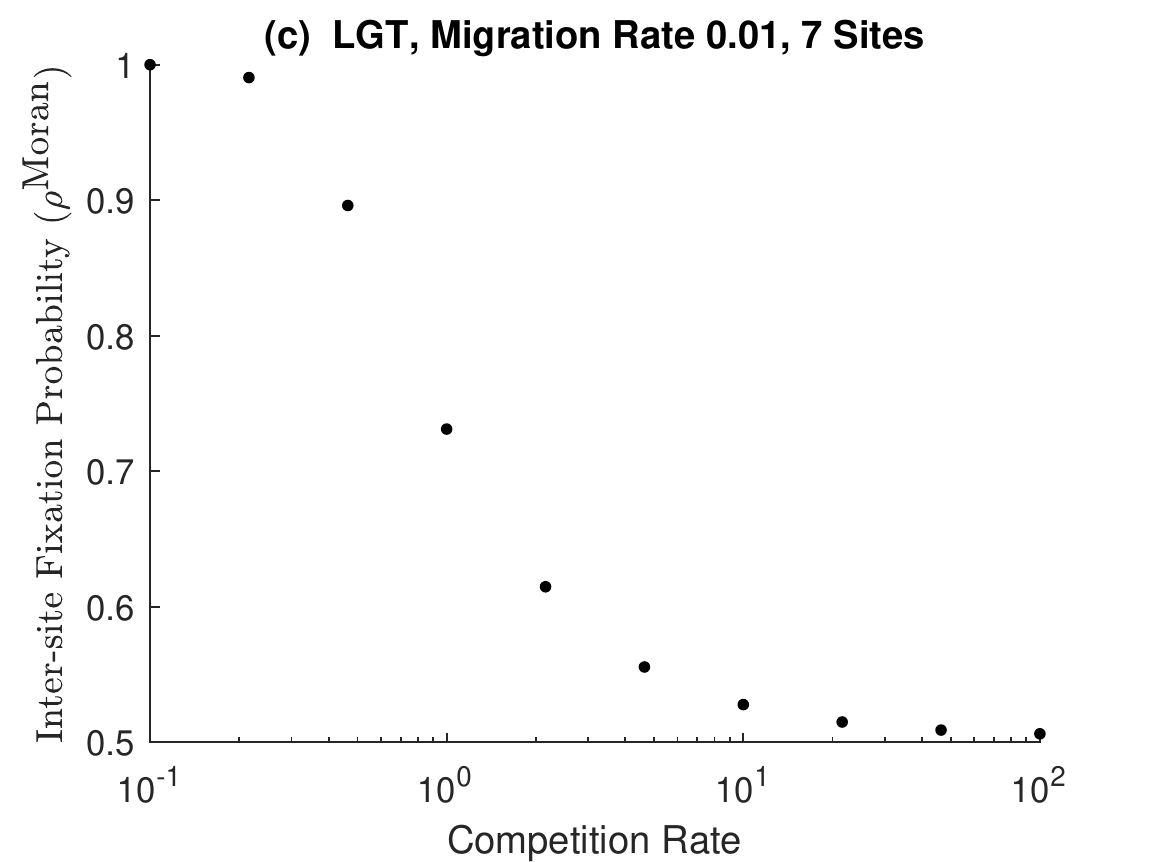}
\includegraphics[width=0.45\linewidth]{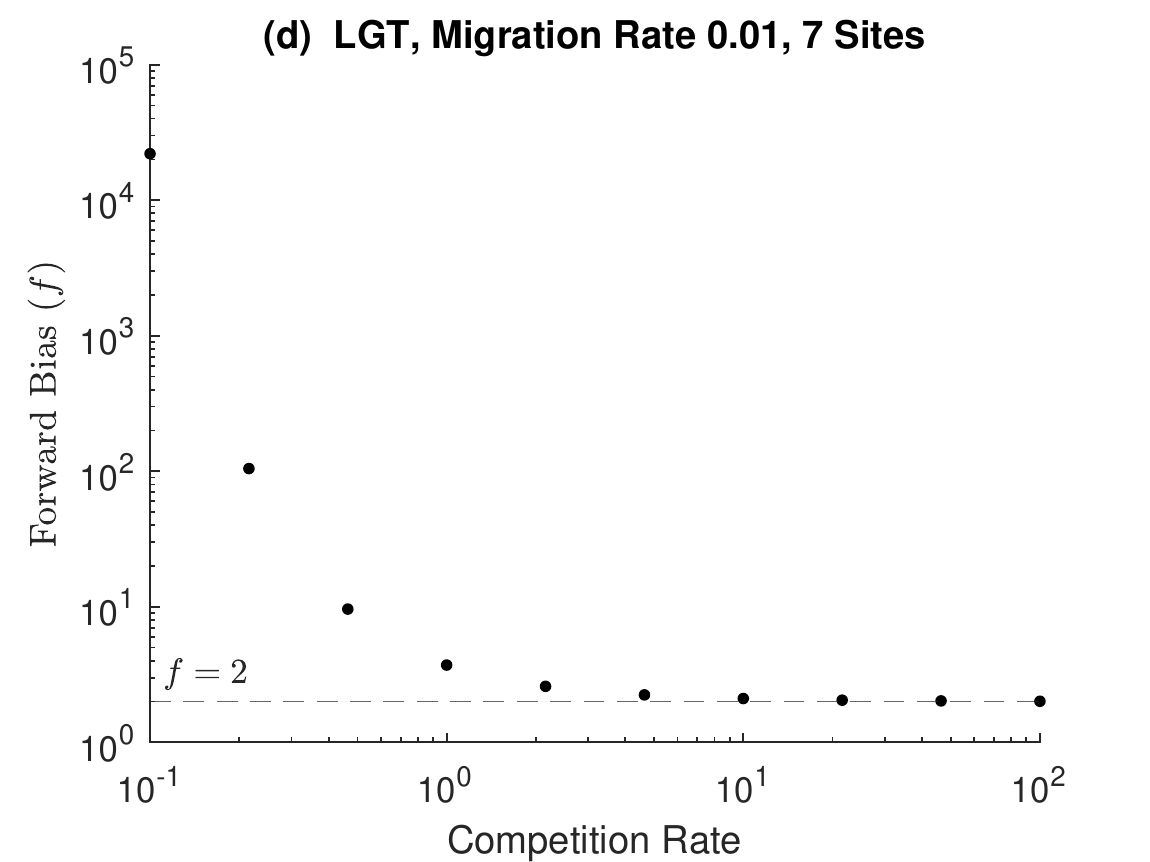}
\caption{
    \label{fig_HGT_lowMove_vsComp}
    Plots for the low migration case.
    (a) Fixation probability of a mutant for low group tolerance (LGT).
    (b) Fixation probability of a mutant for high group tolerance (HGT). 
    In (a--b), `Analytic (LGT)' is analytically calculated by equation \eqref{eq_analytic}.
    In (a), `Anlaytic star (LGT)' is analytically calculated by solving equation \eqref{eq_fixation_low_mig_lgt} using the formula of \cite{broom2008}. 
    (c) Inter-site fixation probability of mutants in circulation networks for low group tolerance, that is, probability of fixating in entire population given that mutants have already fixated in one site.
    (d) The forward bias for mutants in circulation networks for the low group tolerance case.
}
\end{figure}

For low group tolerance, figure \ref{fig_HGT_lowMove_vsComp} (a) shows that an increasing competition rate decreases the fixation probability for a low movement rate in all networks considered.
We use the two components of $\rho^\text{LGT Circ}$ to understand why this is the case.
The first component, $\rho^\text{Single}_M$, describes the intra-site dynamics, which we have already explained.
The second component, $\rho^\text{Moran}$, describes the inter-site dynamics, or how mutants spread once they have fixated on a single site.
The inter-site dynamics are shown in figure \ref{fig_HGT_lowMove_vsComp} (c), where we see that $\rho^\text{Moran}$ is a sigmoid shaped curve whose shape is explained by the forward bias ($f$) that is shown in figure \ref{fig_HGT_lowMove_vsComp} (d).
For a decreasing competition rate we see that
\begin{align}
    \lim_{\gamma\to 0} f\approx \infty \Rightarrow \rho^\text{Moran} \to 1 \Rightarrow \lim_{\gamma\to 0}\rho^\text{LGT Circ}\approx\lim_{\gamma\to0}\rho^\text{Single}.
    \label{eqLowMigLowCompLimit}
\end{align}
This means that for a low competition rate a mutant fixating on one site is sufficient to guarantee that it goes on to fixate in the entire population.
For an increasing competition rate we see that 
\begin{align}
    \lim_{\gamma\to\infty} f& \approx
    \lim_{\gamma\to\infty}
    \frac{
        \frac{(\beta_{M}/\gamma)^2}
        {2!}
        \frac{1}
        {\exp(\beta_M/\gamma)-1}
        \frac{(\beta_{R}/\gamma)^1}
        {1!}
        \frac{1}
        {\exp(\beta_R/\gamma)-1}
        \frac{1}{2}
    }{
        \frac{(\beta_{R}/\gamma)^2}
        {2!}
        \frac{1}
        {\exp(\beta_R/\gamma)-1}
        \frac{(\beta_{M}/\gamma)^1}
        {1!}
        \frac{1}
        {\exp(\beta_M/\gamma)-1}
        \frac{1}{2}
    }
    =
    \frac{\beta_M}{\beta_R}
    \Rightarrow
    \nonumber\\
    \lim_{\gamma\to\infty} \rho^\text{LGT Circ}&\approx \frac{1}{2}\rho^\text{Moran}(N,\beta_{M}/\beta_R) 
\end{align}
where when calculating $f$ we have assumed that there are two individuals when a migration event happens and that an immigrant arrives to a site with one individual only so fixates with probability $\frac{1}{2}$.
Note that in this case the forward bias converges to the relative birth rates of the individuals ($\beta_M/\beta_R=2$), which means that each site can be viewed as a single individual as in the case of EGT.
The inter-site dynamics are therefore equivalent to Birth-death (Bd) dynamics where an individual is selected proportional to their fitness to replace a randomly chosen individual such that fitness is proportional to the birth rates of individuals \citep{pattni2021}.

In the star network for low group tolerance figure \ref{fig_HGT_lowMove_vsComp} (a) shows that it follows a similar pattern to the complete and cycle networks.
However, the fixation probability in the star network is higher for high competition rates but converges as the competition rate decreases. 
Since $\lambda\to0$, mutants are likely to appear on leaf sites in a star network as the combined number of individuals on leaf sites is higher than the centre site.
Appearing on leaf sites is beneficial because the way in which $W$ is defined for the star network (equation \eqref{eq:weights}) allows leaf sites to act as source sites, i.e.~are net exporters of individuals, \citep{pattni2021}.
For low competition rate, convergence occurs since the intra-site dynamics are identical for all networks and, as explained earlier,  if a mutant fixates on one site, it is essentially guaranteed to fixate in the entire population.
On the other hand, as the competition rate increases, which gives residents a better chance to prevent invasion, the divergence in the fixation probability between the star and circulation networks becomes more apparent. 

For high group tolerance, figures \ref{fig_HGT_lowMove_vsComp} (b) shows a similar pattern to low group tolerance where fixation probability decreases as the competition rate increases.
High group tolerance allows empty sites, however, when competition rate is low, the likelihood of empty sites deceases and the intra-site dynamics would be similar to that of low group tolerance.
The fixation probabilities are therefore identical to the low group tolerance case for low competition rate.
As the competition rate increases, the chance of having empty sites increases, changing the behaviour observed.
In particular, the population starts converging to a population size of 1 as individuals start dying off when they meet.
This means that as the competition rate increases the fixation probability starts converging to $\frac{1}{2}$ as the likelihood that a mutant appears in a population with one resident increases.
Overall, the fixation probability start decreasing then increasing again as the competition rate increases.

\newpage

\newpage
\subsection{General Migration Rate}
 In this section we consider the case for a general migration rate ($\lambda>0$). 
The fixation probability in this case is calculated via simulation.

\subsubsection{Effect of increasing migration rate}
Migration allows individuals to escape competition as shown in figure \ref{fig_LGT_movement} where the fixation probability increases with the migration rate.
The way in which this plays out depends upon the competition rate, network structure and group tolerance.
The effect of these are explained in the following.

\begin{figure}
    \centering
    \includegraphics[width=0.5\textwidth]{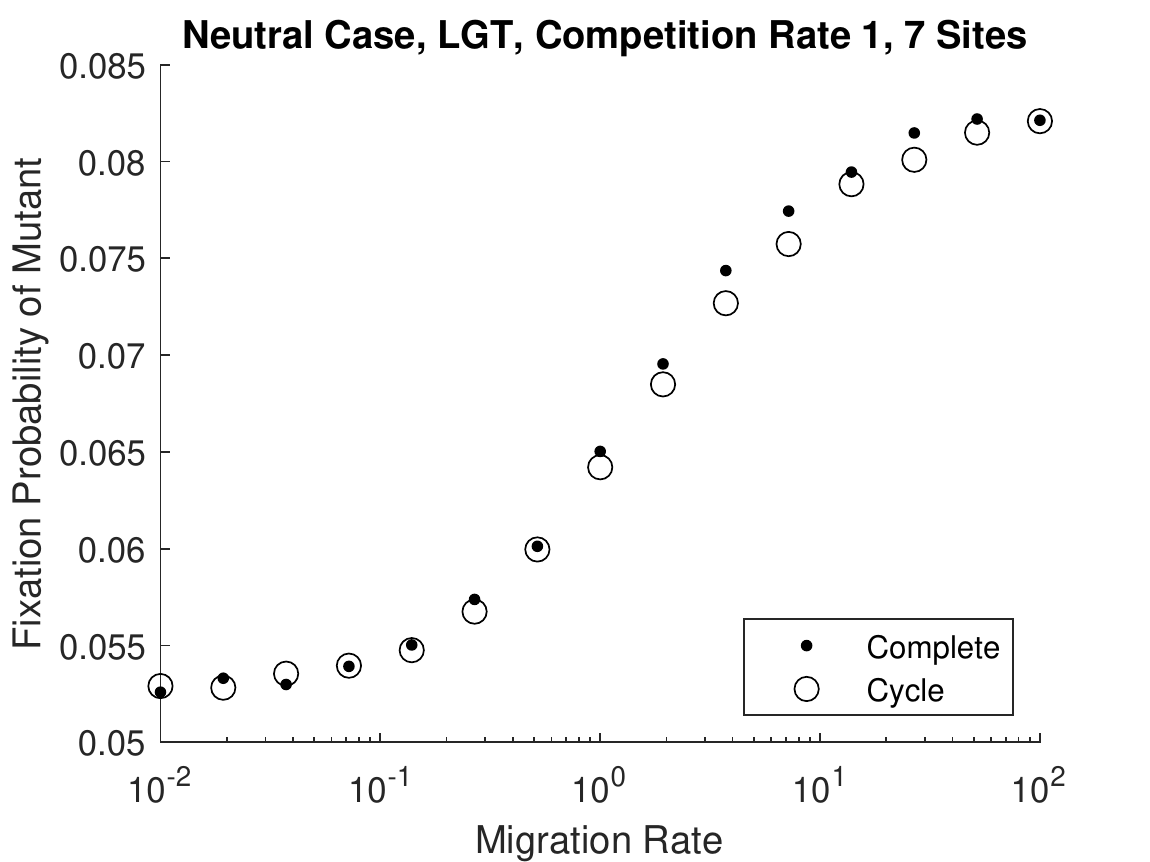}
    \caption{Fixation probability in the neutral case ($\beta_R=\beta_M=1$). This plot was generated using $10^6$ simulations.}
    \label{fig:movement_neutral}
\end{figure}

For low group tolerance, figure \ref{fig_LGT_movement} (a--d) shows that as the migration rate increases the fixation probability starts increasing and plateaus earlier for low competition than high competition.
However, as $\lambda\to\infty$ there would be a larger overall increase in the fixation probability for high competition.
In the initial growth and plateau phases of the fixation probability, the complete and cycle networks follow each other closely and are indistinguishable.
As the growth in the fixation probability accelerates, there is higher acceleration in the complete network than the cycle network.  
The key factor here is local correlation between groups on neighbouring sites on the cycle.
For low migration rates fixation probabilities are low and similar for both cycle and complete networks.
New individuals are likely to be born into bigger groups, as there are more potential parents, but cannot move on, so face increased competition. 
As they hardly move, network does not matter.
For intermediate migration rates fixation probabilities are intermediate, but differ for the two types.
New individuals are born to bigger groups, but there is some dispersal so they face an intermediate level of competition. 
Here dispersal happens to some extent, and so the network does matter.
For high migration rates fixation probabilities are high and similar for the two types.
New individuals are born in bigger groups but then there is rapid dispersal so they live in `average' groups. 
As migration is so far they mix well, so network does not matter.
To illustrate this point further, figure \ref{fig:movement_neutral} shows the fixation probability in the case of a neutral mutant, i.e.~$\beta_R=\beta_M=1$. 
If there is no correlation between the sites, the fixation probability would be identical for the complete and cycle networks.
There is correlation as we see a difference in the fixation probabilities, which happens for intermediate migration rates.
For the star network, as $\lambda$ increases we see that there is an initial dip in the fixation probability before it starts increasing.
This is because increasing the migration rate results in the number of individuals in the centre site becoming larger than a leaf site.
This increases the likelihood of a mutant appearing in the centre site which is a sink, i.e.~a net importer of individuals, which adversely affects the fixation probability \cite{pattni2021}.
This dip happens earlier for a lower competition rate and, after this dip, the fixation probability remains below that of the complete and cycle networks.

For high group tolerance figure \ref{fig_LGT_movement} (e--h) shows that the behaviour observed is similar to to low group tolerance when competition rate is low but vastly different for a higher competition rate.
For low competition rate, the intra-site dynamics are similar in high and low group tolerance.
In particular, for a low competition rate the likelihood of there being empty sites is low even as the migration rate increases.
On the other hand, for a high competition rate the likelihood of empty sites increases.
This means that a mutant arises in a population with fewer individuals than in the low group tolerance case.
This is observed in figure \ref{fig_LGT_movement} (g) and (h). 
In (g), the star network has a higher fixation probability for all migration rates than the complete and cycle networks.
This is because individuals are more likely to meet in the centre site resulting in death due to competition, which drives the population size down.
This effect is substantial for a high competition rate as seen in (h). 
As the migration rate increases, the fixation probability in all networks swiftly converges to $\frac{1}{2}$ since the population size is converging to 1.

\begin{figure}
\centering
    \includegraphics[width=0.4\linewidth]{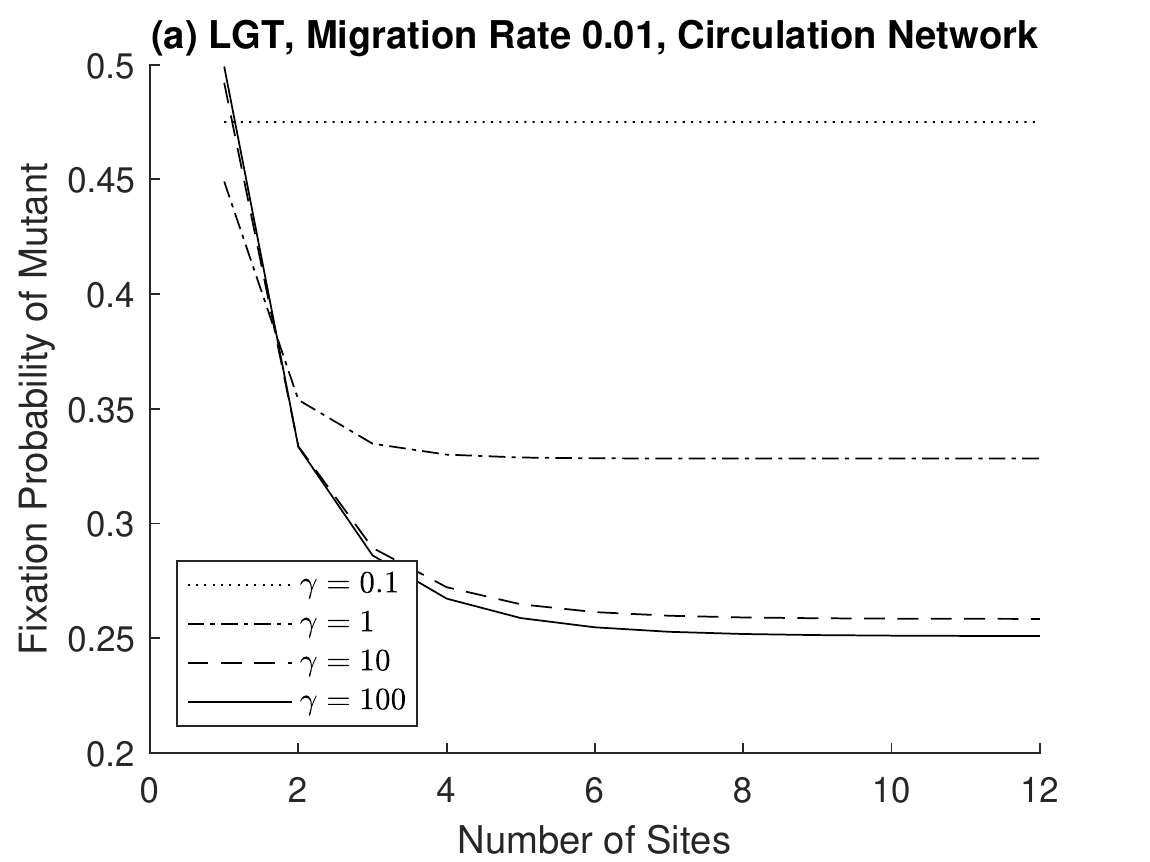}
    \includegraphics[width=0.4\linewidth]{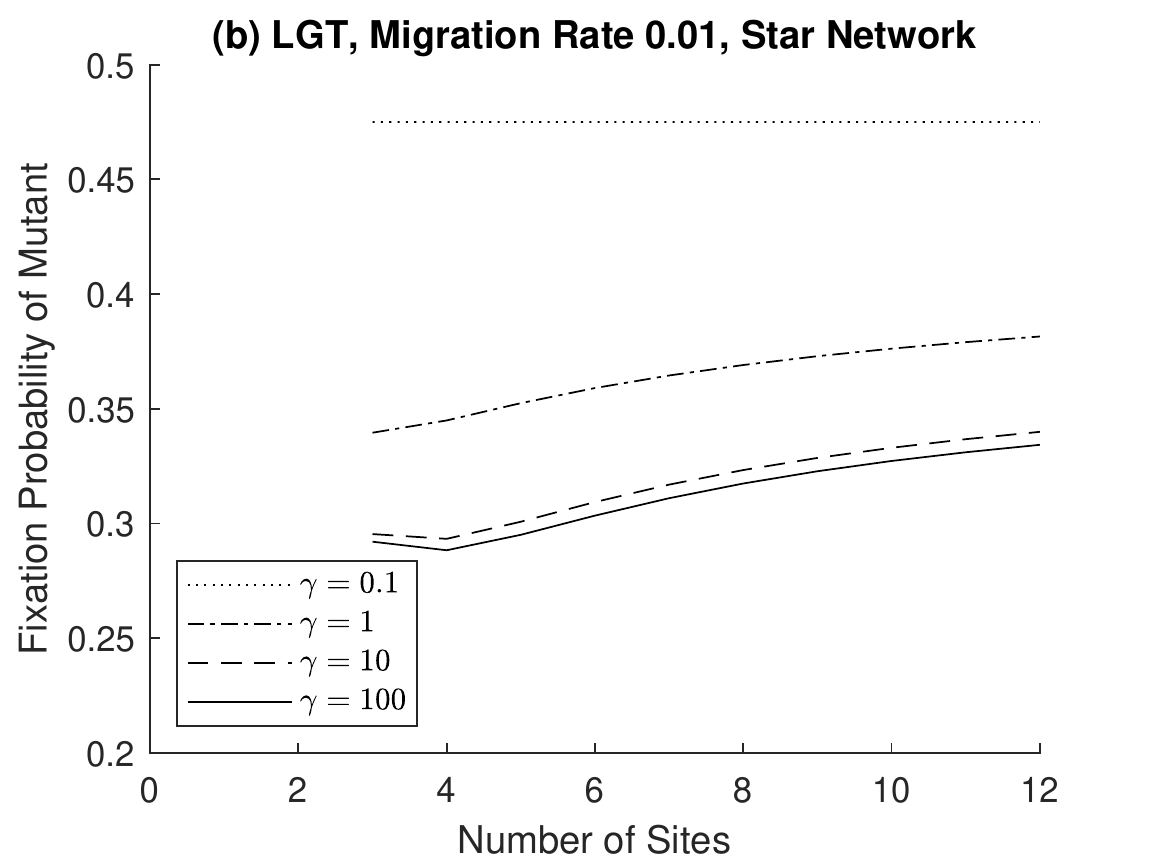}
    \centering
    \includegraphics[width=0.4\linewidth]{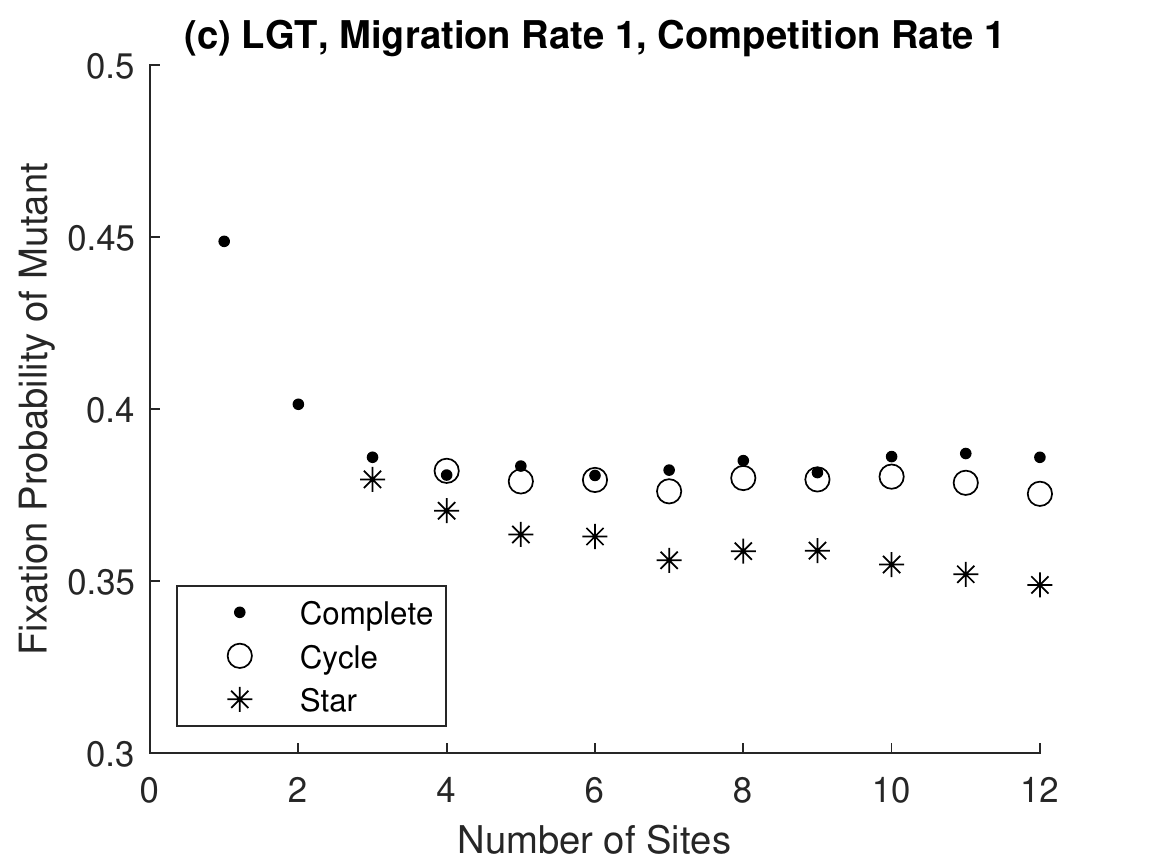}
    \includegraphics[width=0.4\linewidth]{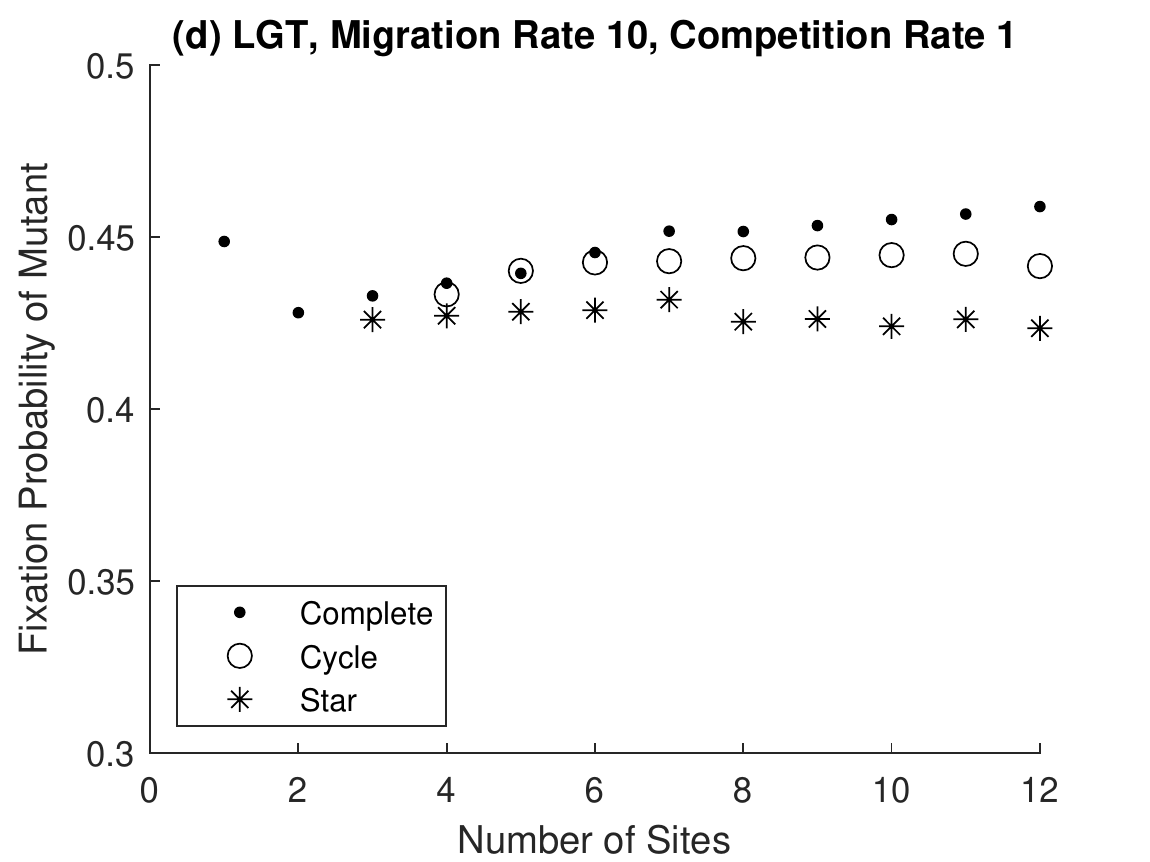}
    \includegraphics[width=0.4\linewidth]{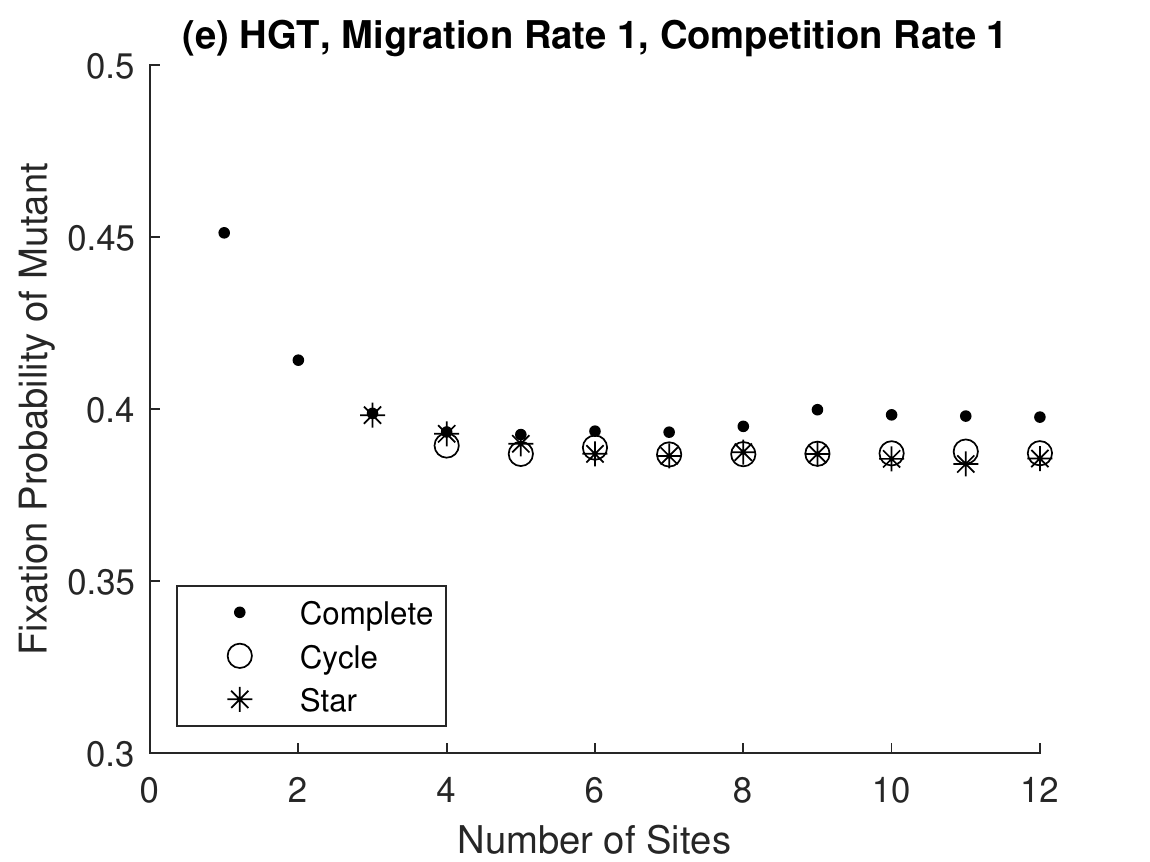}
    \includegraphics[width=0.4\linewidth]{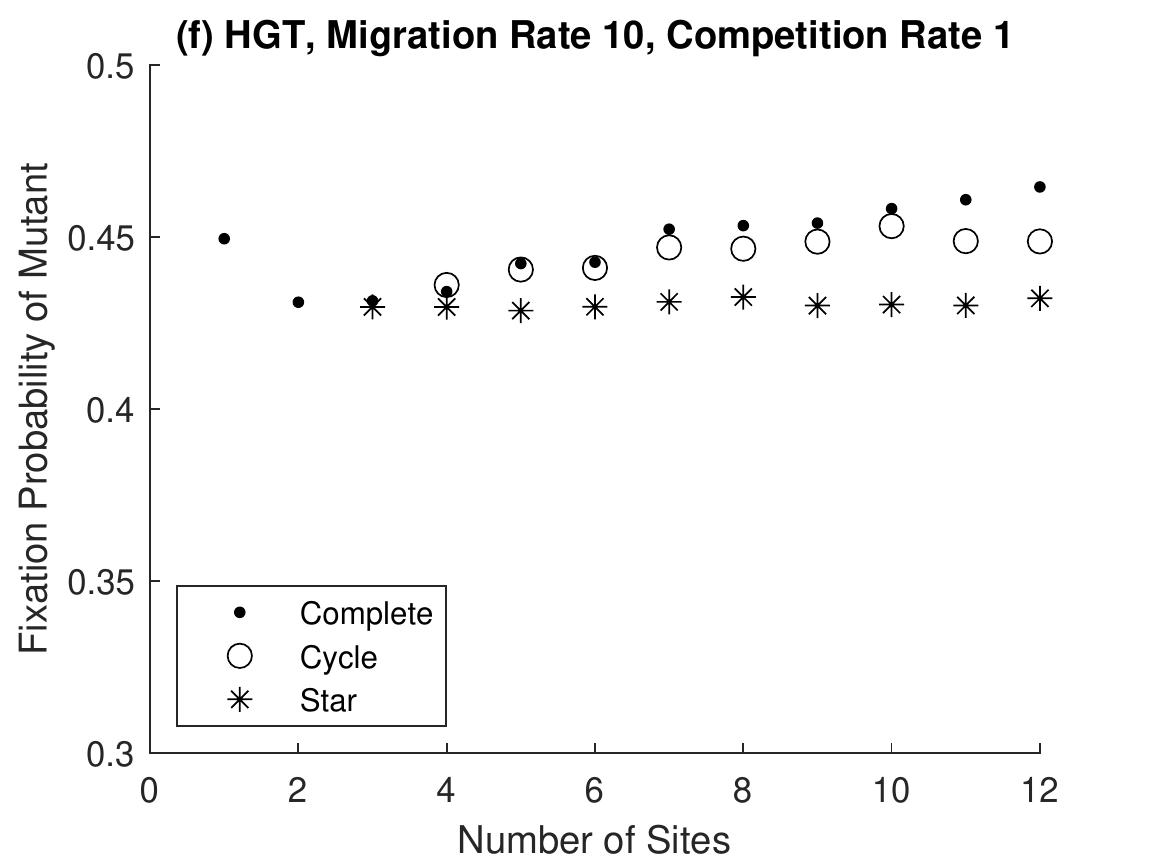}
    \caption{Effect of increasing the number of sites on the fixation probability. 
    Figures (a--d) are for low group tolerance (LGT) and figures (e--f) are for high group tolerance (HGT).
    Figure (a) is analytically calculated using equation \eqref{eq_analytic} for Circulation networks, and figure (b) is calculated using the analytical formula for the star network.
    Figures (c--f) are generated using $10^5$ simulations.
    For the cycle network, the plot starts from 4 sites as fewer than 4 sites classifies as a complete network.
    For the star network, the plot starts from 3 sites as this is the minimum number of sites required to construct a star network.
    \label{fig:n_sites}
    }
\end{figure}

Figure \ref{fig:n_sites} shows the effect of migration rate as the number of sites increases.
Figure \ref{fig:n_sites}(a) considers the low migration limit for low group tolerance in circulation networks.
We see that for a low competition rate ($\gamma=0.1$), the fixation probability remains the same as the number of sites increases.
This was previously explained using equation \eqref{eqLowMigLowCompLimit}, where fixating in one site was sufficient to guarantee fixation in all sites.
This effect carries over for higher competition rates, but the number of sites required to guarantee fixation increases.
We observe that the fixation probability initially starts to decrease as the number of sites increases, but once we reach the point where the number of sites guarantees fixation, the fixation probability will flatline.
For example, we see that for a high competition rate ($\gamma=100$)  the fixation probability flatlines after approximately 7 sites, that is, fixating in 7 sites guarantees fixation in the entire population. 
This effect is also evident in circulation networks for low group tolerance and high group tolerance with a relatively low migration rate of $1$ as seen in figures \ref{fig:n_sites} (c) and (e). 
However, when the migration rate is increased ($\lambda = 10$), for both low and high group tolerance there is a slight dip as the number of sites increases, but recovers to the one site level as seen in figures \ref{fig:n_sites} (d) and (f).
This is because the population effectively behaves as one big unit when the migration rate is high, with this being more pronounced with a higher number of sites as there are more individuals.
In the star network (figure \ref{fig:n_sites} (b)), for low competition the effect of increasing the number of sites is the same as for circulation networks i.e.~fixating in one site guarantees fixation in the entire population.
For a higher competition rate, the fixation probability increases with the number of sites. 
This is because we are adding a leaf site each time the number of sites increases, which increases the likelihood of a mutant appearing on a leaf site.
As explained earlier, leaf sites are source sites and therefore beneficial for a mutant.
As the migration rate increases to 1, the fixation probability decreases in the star network with an increasing number of sites for low group tolerance (figure \ref{fig:n_sites} (c)).
This is because the higher migration rate results in an increased number of individuals in the centre site, which increases the likelihood of the initial mutant appearing in the centre site. 
As the centre site is a sink, it is less beneficial for mutants.
This does not happen for high group tolerance (figure \ref{fig:n_sites} (e)), as most leaf sites are likely to remain empty with most of the population being present in the centre site.
The population therefore behaves as one large unit clustered in the centre site.
As the migration is increased to 10, for both low and high group tolerance (figure \ref{fig:n_sites} (d) and (f)), the fixation probability in the star network remains constant as the number of sites is increased.
This is because individuals are mixing with each other much more, nullifying the effect of network structure in both cases.

\noindent
\begin{minipage}{\linewidth}
    \centering
    \captionsetup{type=figure}
    \includegraphics[width=0.4\linewidth]{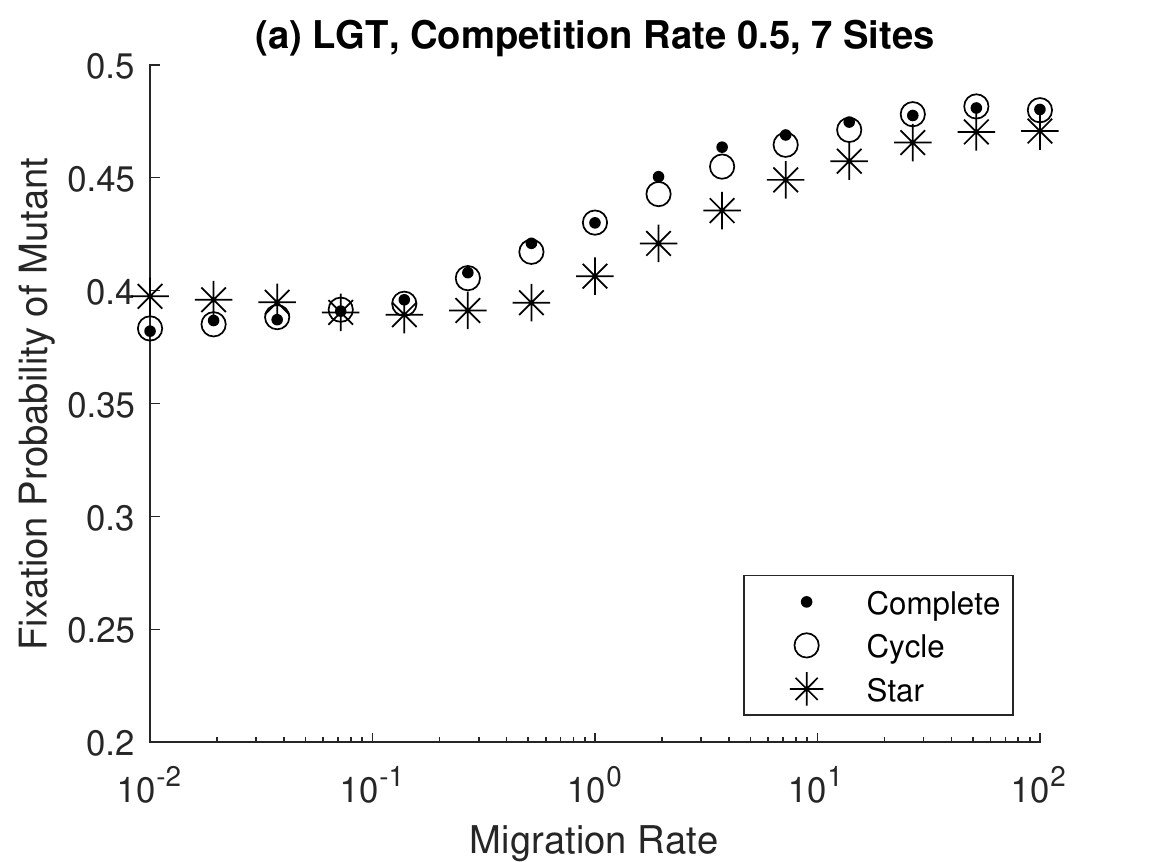}
    \includegraphics[width=0.4\linewidth]{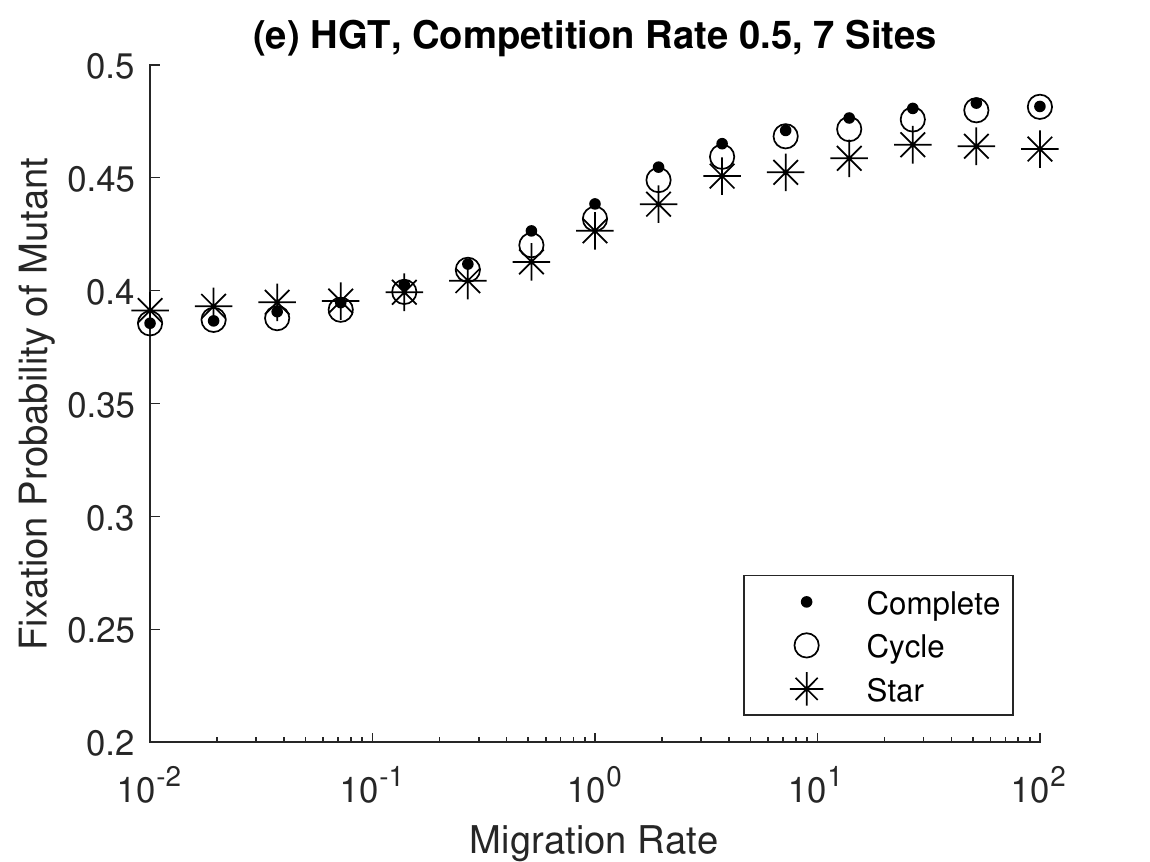}
    \includegraphics[width=0.4\linewidth]{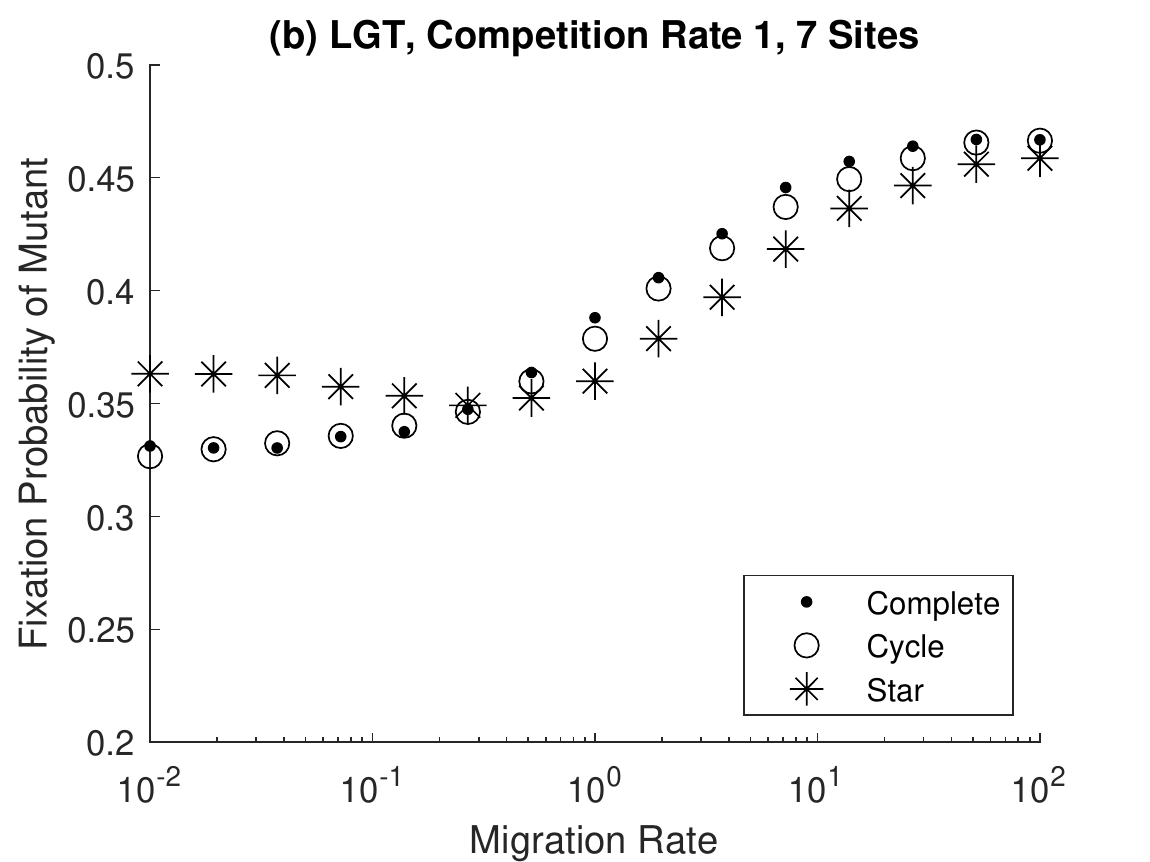} \includegraphics[width=0.4\linewidth]{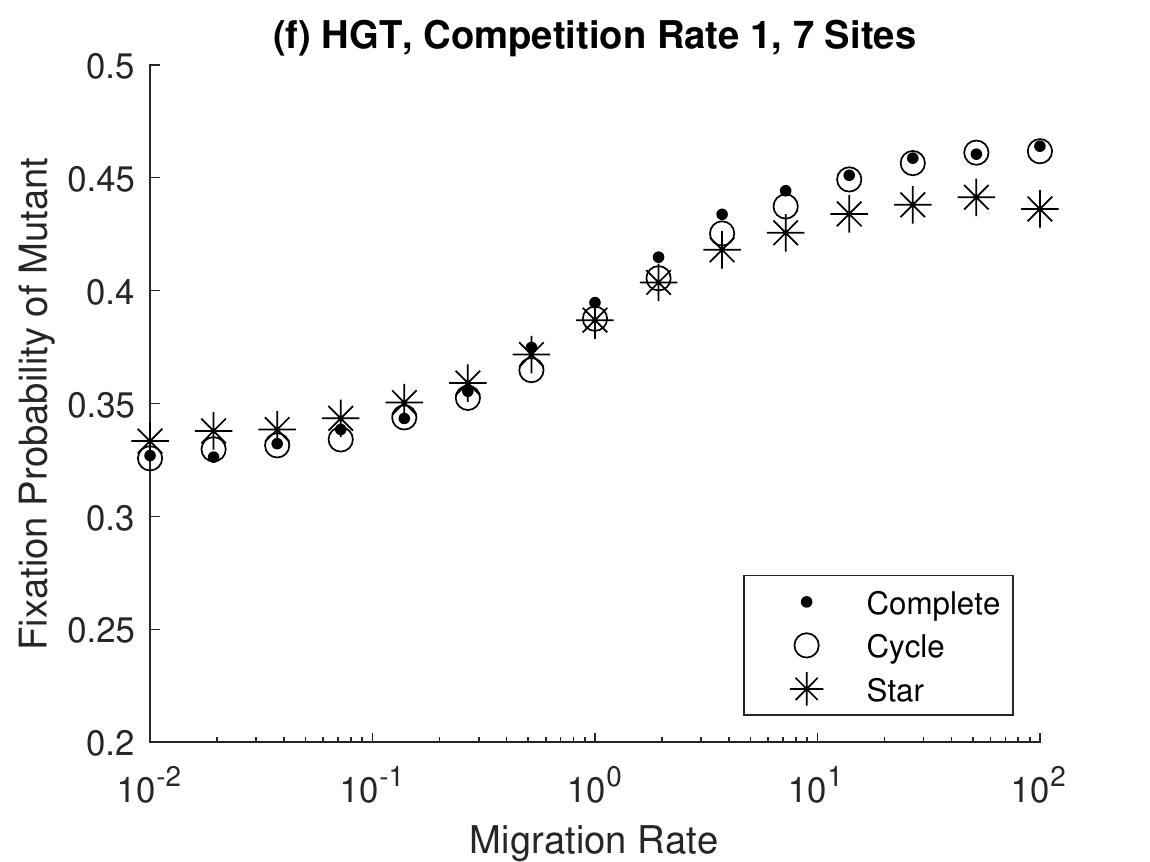}
    \includegraphics[width=0.4\linewidth]{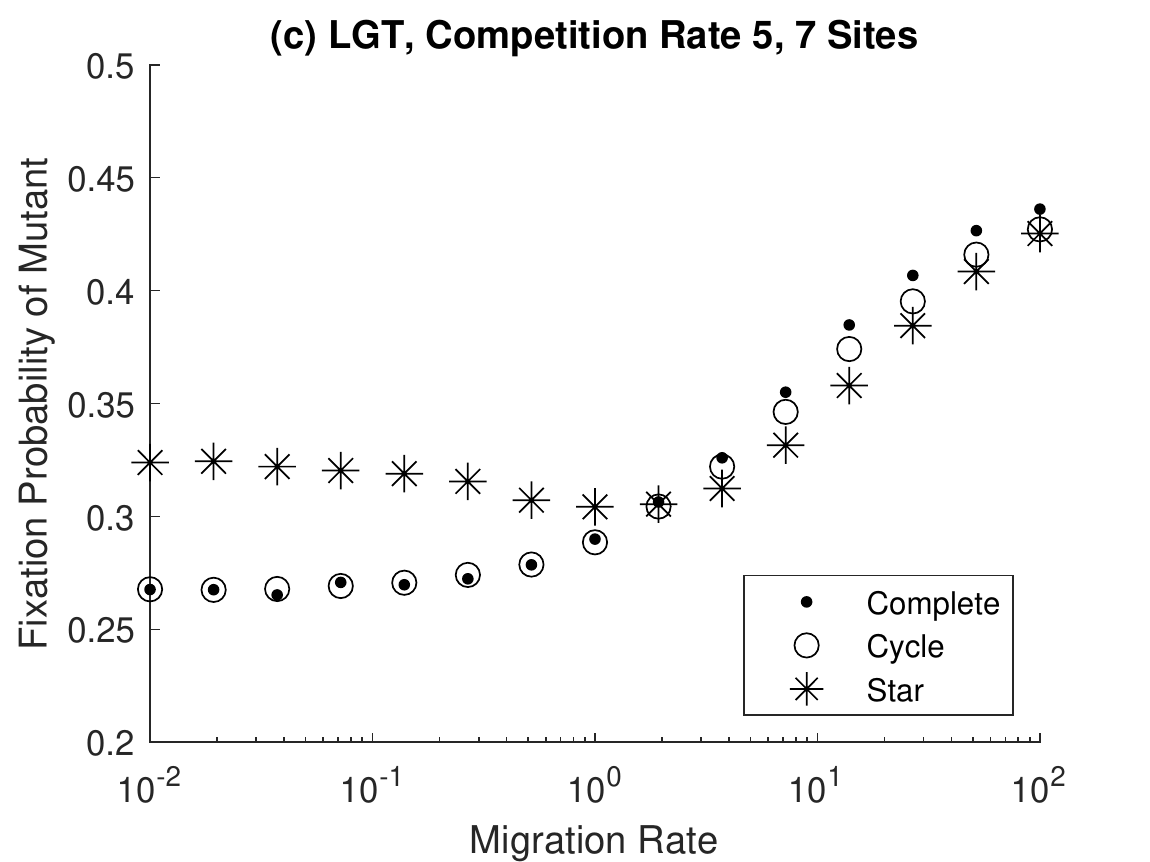}
    \includegraphics[width=0.4\linewidth]{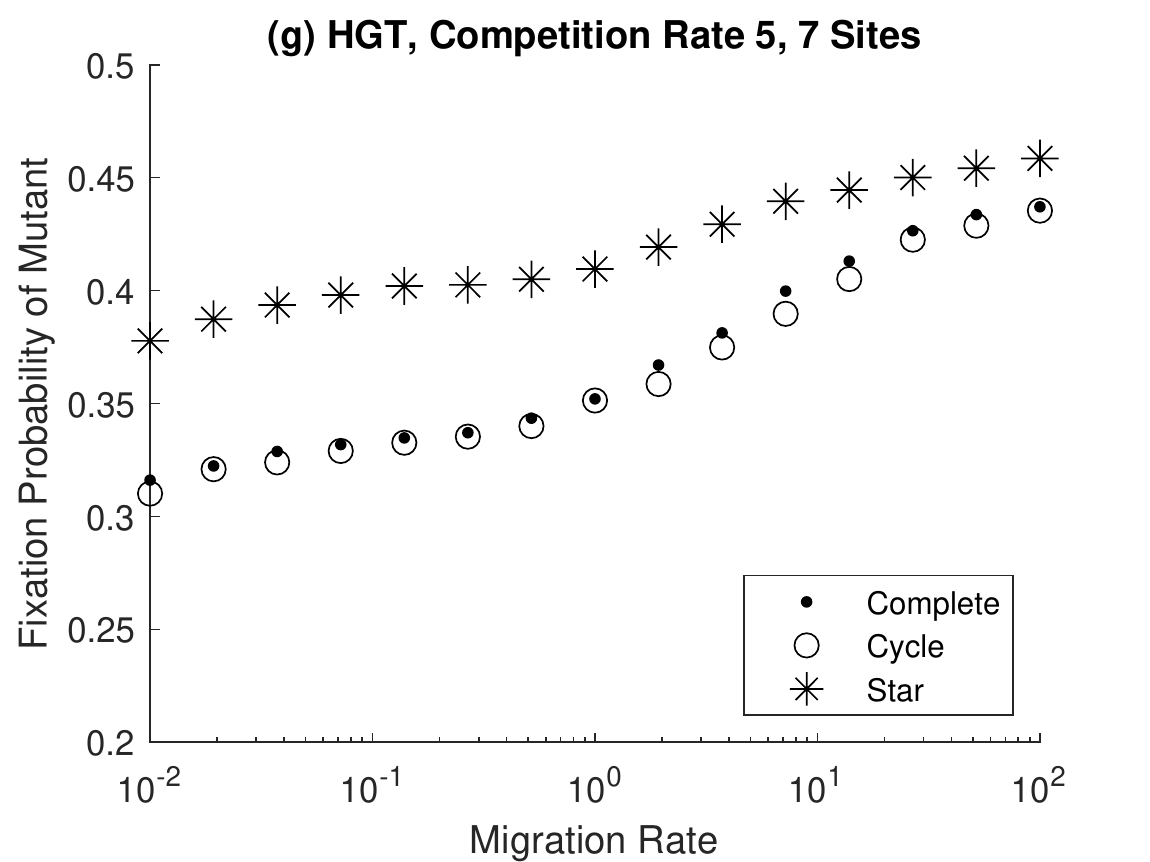}
    \includegraphics[width=0.4\linewidth]{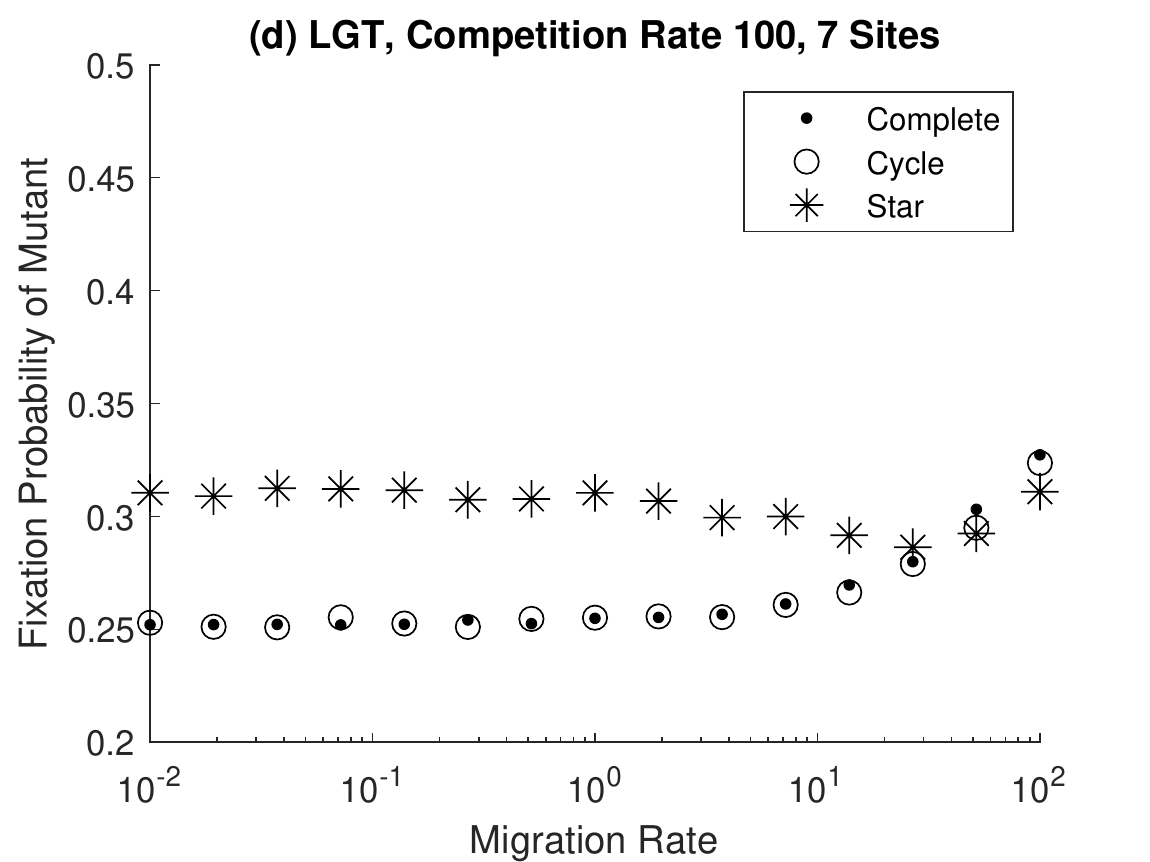} 
    \includegraphics[width=0.4\linewidth]{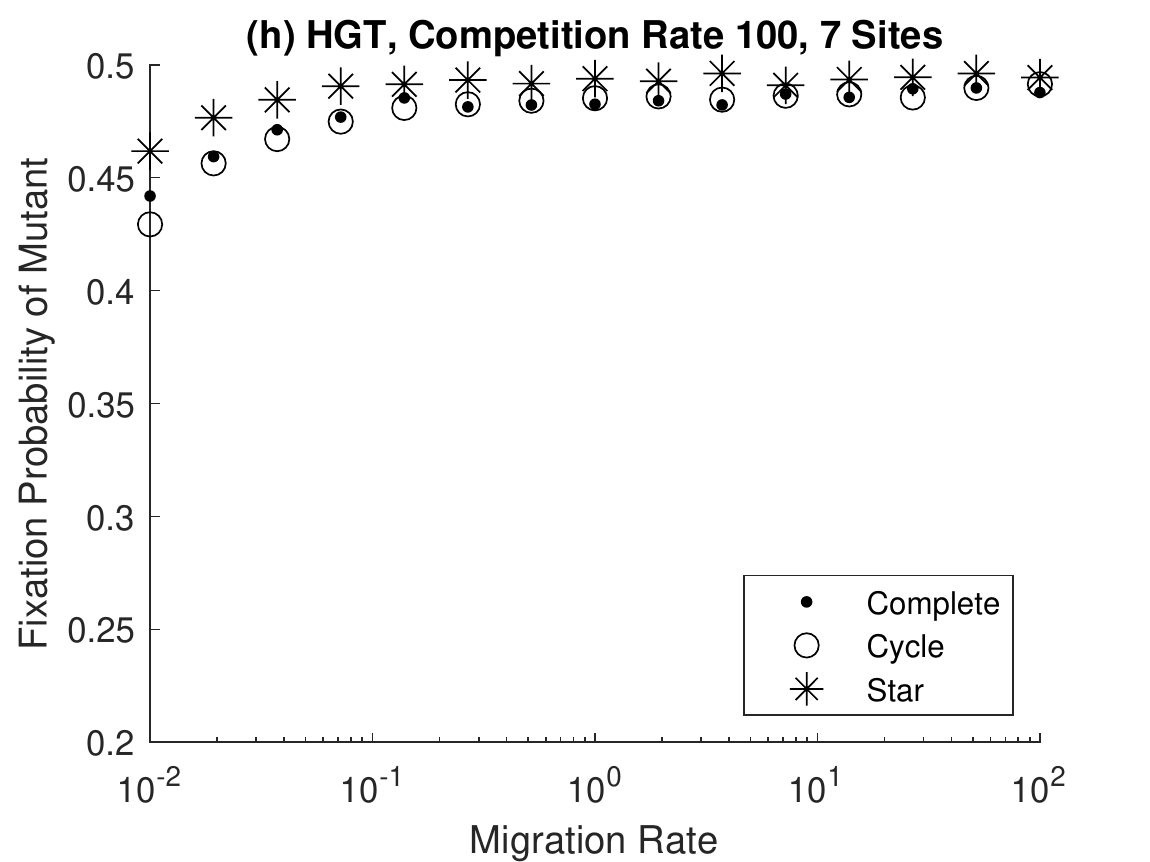}
    \captionof{figure}{Fixation probability of a mutant plotted against the migration rate of individuals for different competition rates. 
    Each network has 7 sites and birth rate of mutants is 2. Figures (a--d) there is low group tolerance (LGT). Figures (e--g) there is high group tolerance (HGT).}
    \label{fig_LGT_movement}
\end{minipage}

\subsubsection{Comparison to Low Migration Limit}

For low group tolerance Figure \ref{fig_LGT_competition} (a--c) shows that there is initially similar behaviour to the low migration limit, but gradually breaks down as the migration rate keeps increasing.
The fixation probability in the complete and cycle networks are higher than in the low migration limit as migration enables escaping competition.
This difference is less apparent for high competition as it requires a much larger migration rate to make a significant difference.
In the low migration limit, the star network has a higher fixation probability than the complete and cycle networks, and converges as the competition rate decreases. 
Here, the star network initially has a lower fixation probability for a low competition rate.
As the competition rate increases this difference gradually diminishes and eventually the fixation probability surpasses that of the complete and cycle networks. 
This behaviour is explained by the decreasing likelihood of a mutant appearing in the centre site.
When the competition rate is low, there are more individuals in the centre site than there are on a leaf site, thus there is an increased likelihood of a mutant appearing the centre site.
However, this likelihood starts decreasing as the competition rate increases, which reduces the number of individuals in the centre site.
Since the centre site acts as a sink, i.e.~it is a net importer of individuals \citep{broom2008,pattni2021},
it suppresses the fixation probability.

For high group tolerance figure \ref{fig_LGT_competition} (d--f) shows that the fixation probability decreases then increases as the competition rate increases. 
This behaviour is significantly different to that observed in the low migration limit.
For low competition rate ($\gamma \le 1$) the behaviour is similar to that of low group tolerance (figure \ref{fig_LGT_competition} (a--c)), this means that the intra-site dynamics are similar for both cases.
That is, even though high group tolerance allows for empty sites in their intra-site dynamics, this is unlikely when the competition rate is lower.
As the competition increases ($\gamma >1$), empty sites are more likely as a death is more likely to occur whenever moving individuals come into contact with one another.
This drives the population size down.
We observe that as the competition rate increases, the fixation probability turns and starts to increase, eventually converging to 0.5.
This implies that the entire resident population prior to a mutant arising is converging to 1.
When comparing between networks, the fixation probability in the star network turns first and converges faster to 0.5. 
This is because individuals are more likely to meet in the center site in a star network which drives the population size down faster than the complete and star networks.

\noindent
\begin{minipage}{\linewidth}
\captionsetup{type=figure}
\includegraphics[width=0.5\textwidth]{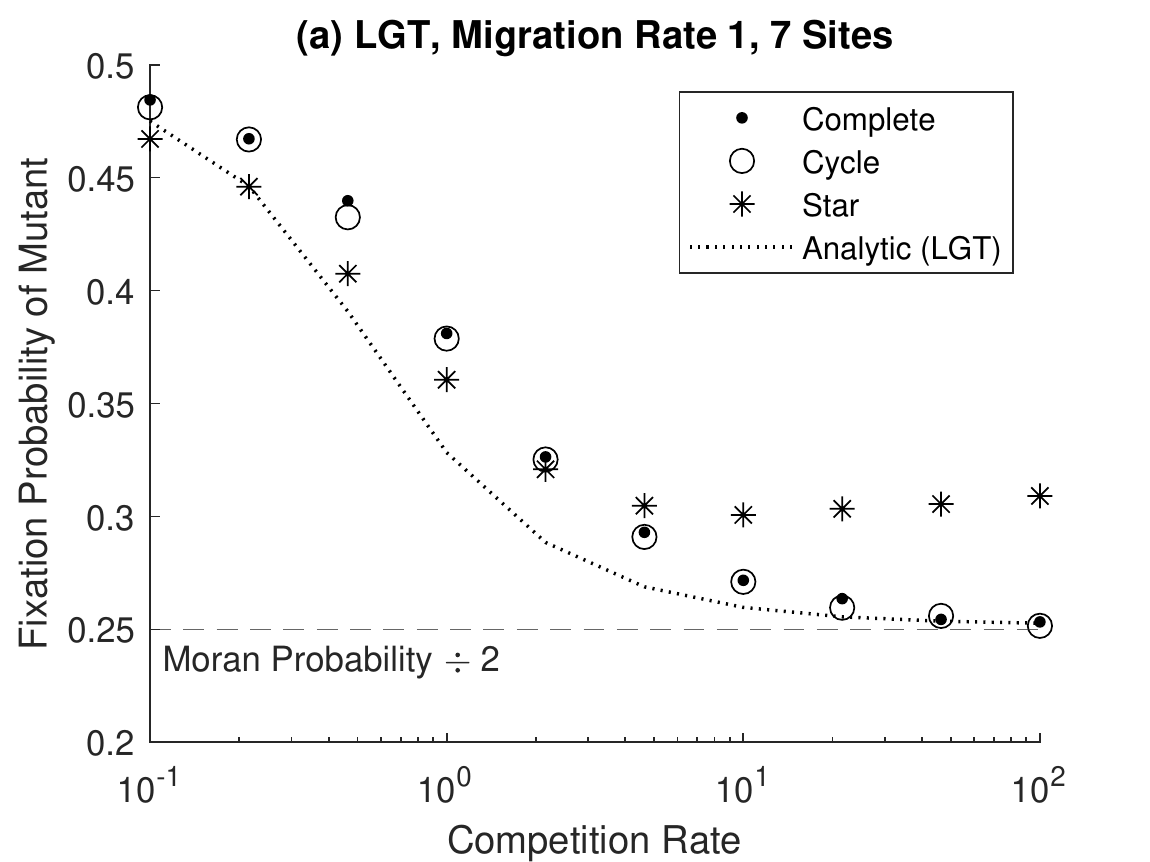}
\includegraphics[width=0.5\textwidth]{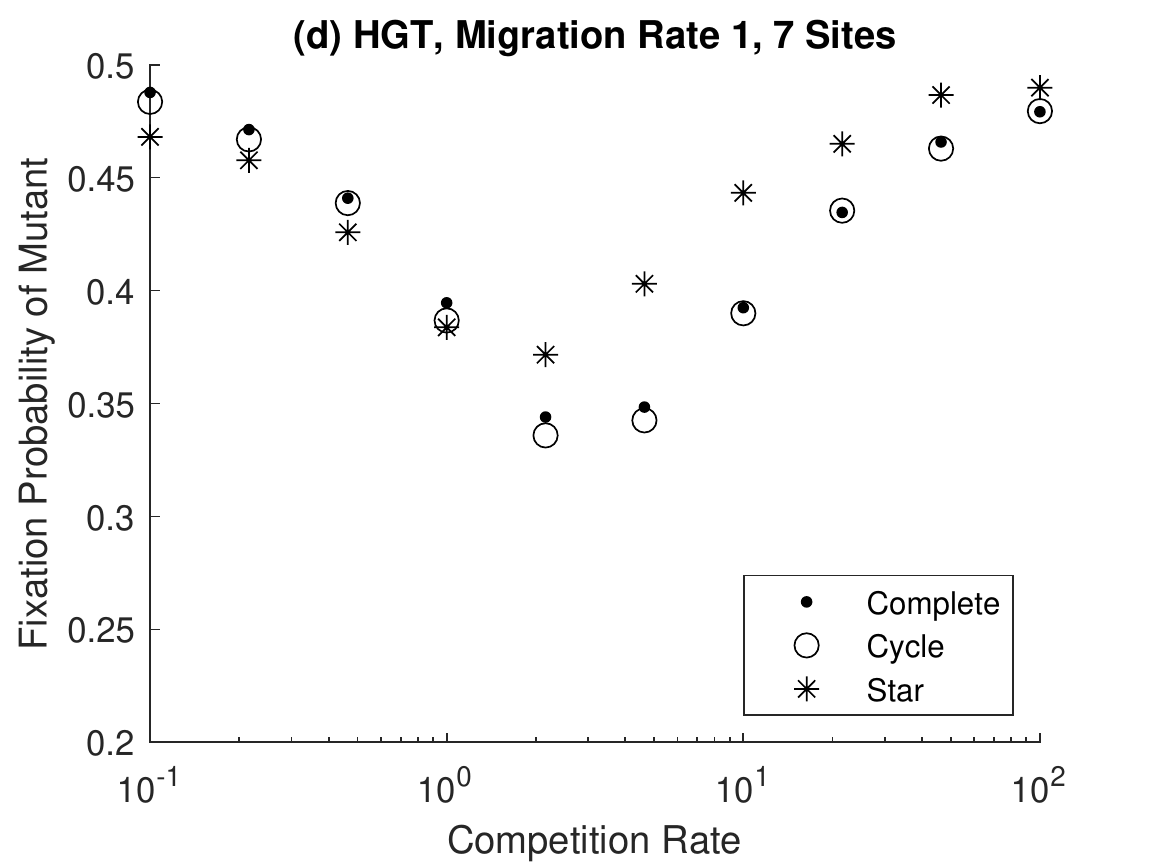}
\includegraphics[width=0.5\textwidth]{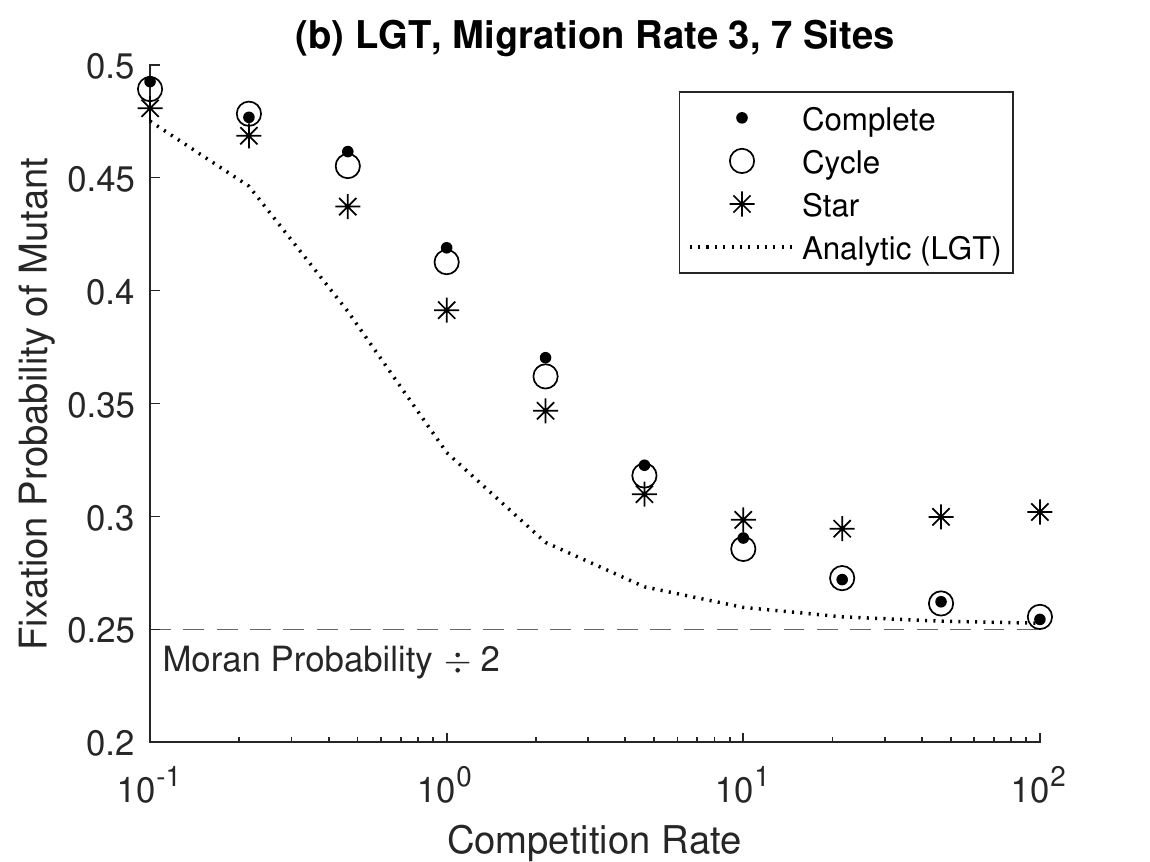}
\includegraphics[width=0.5\textwidth]{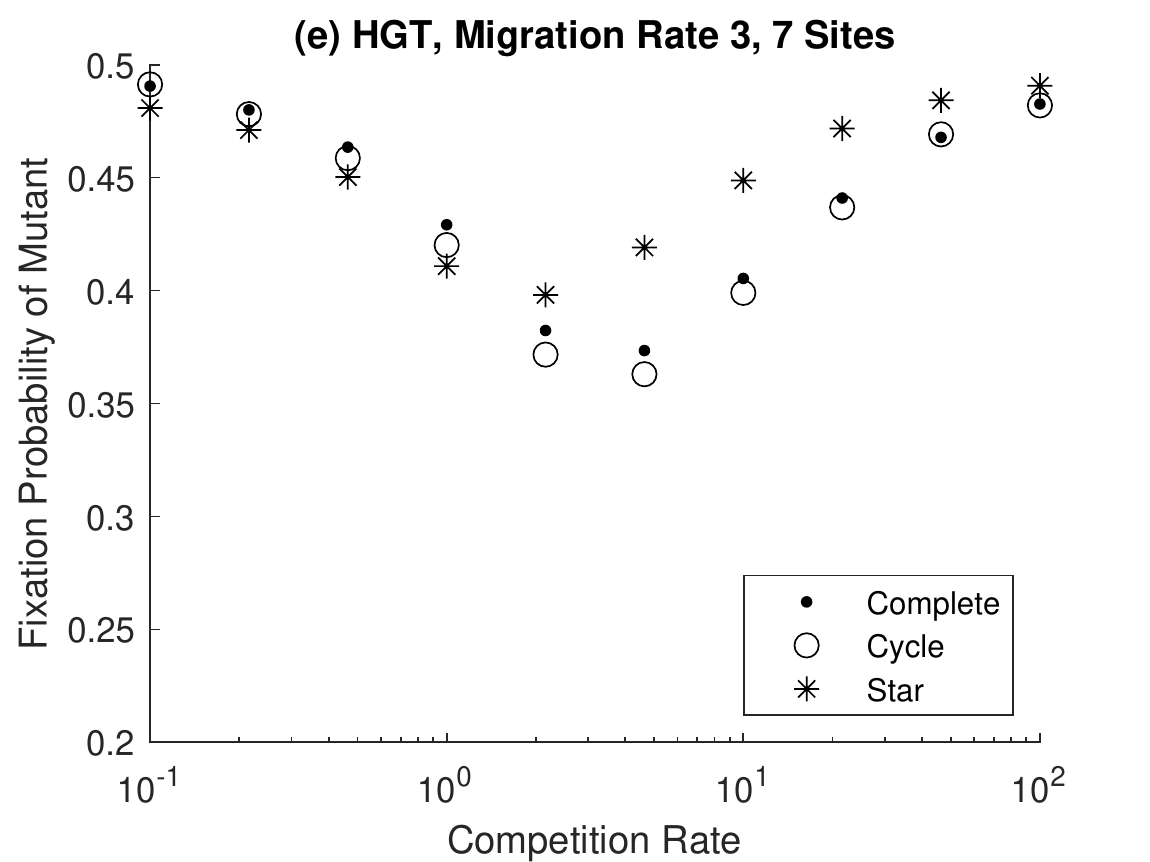}
\includegraphics[width=0.5\textwidth]{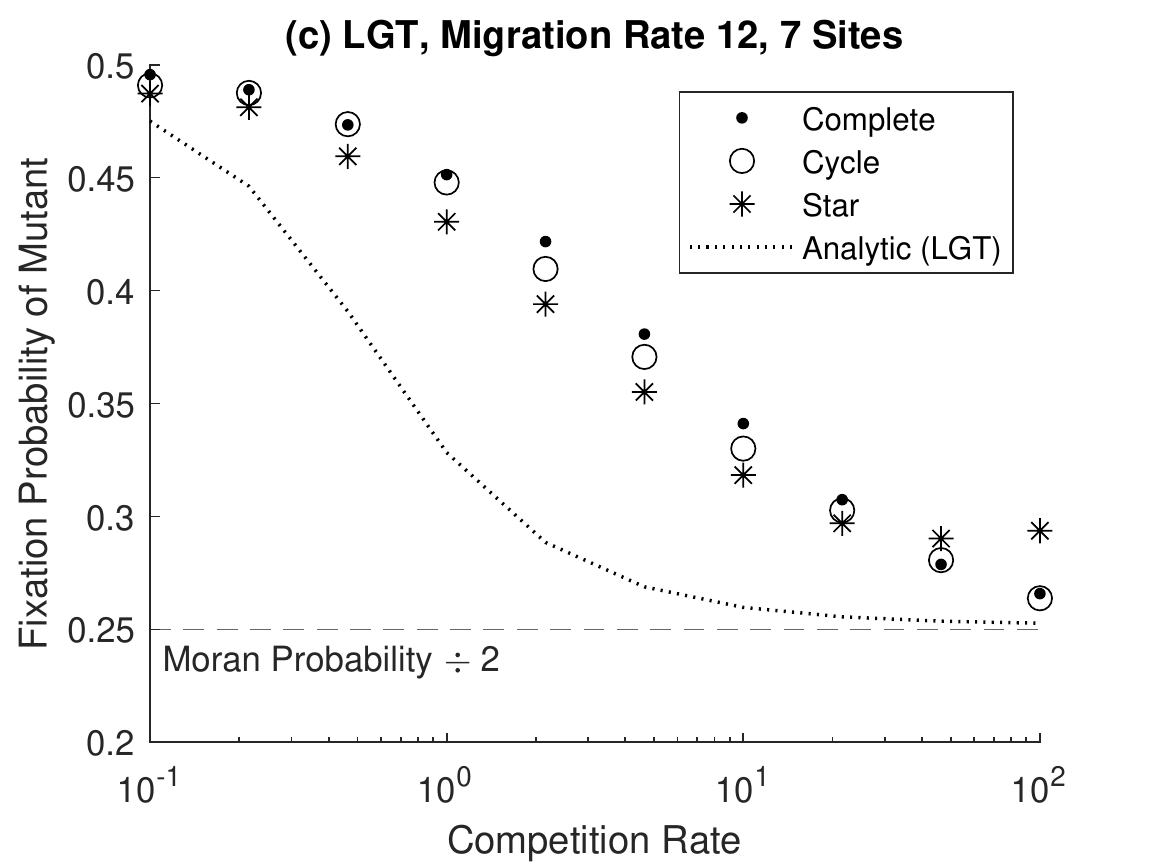}
\includegraphics[width=0.5\textwidth]{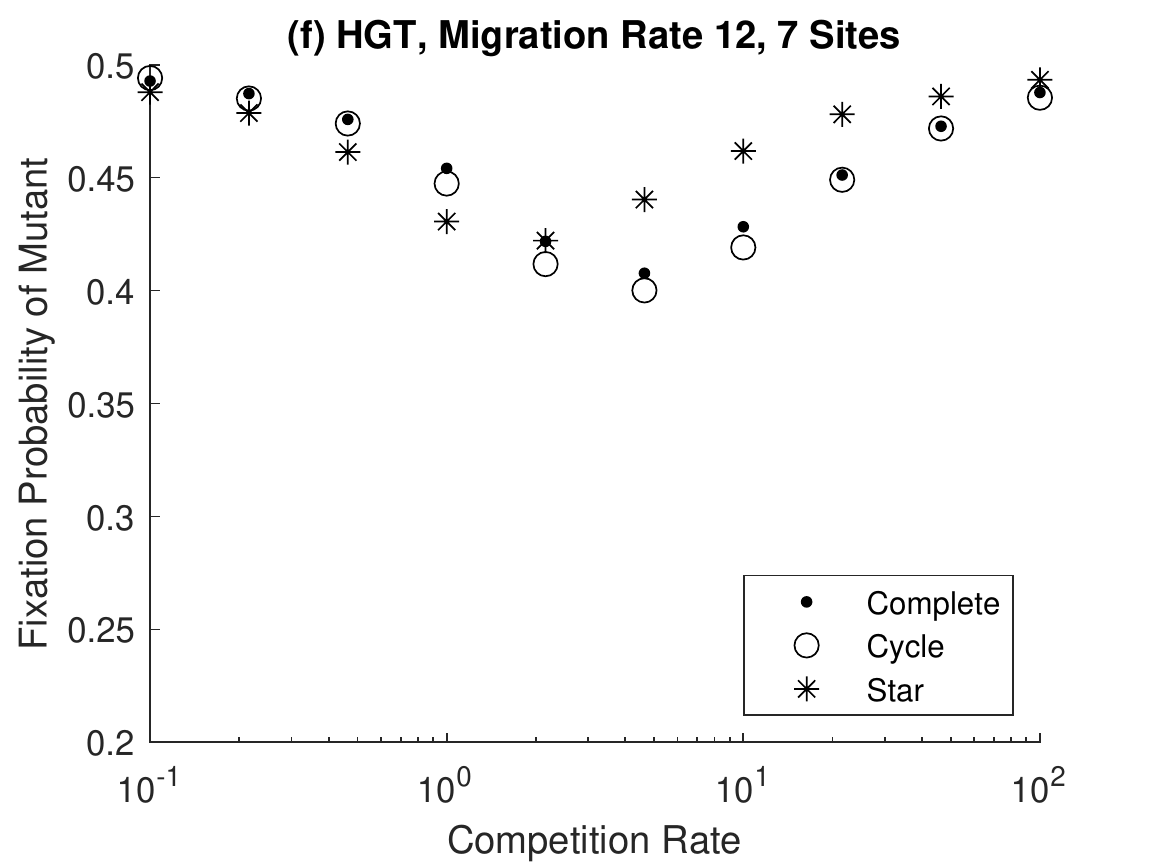}
\captionof{figure}{
\label{fig_LGT_competition}
Fixation probability in the cycle, complete and star networks plotted against the competition rate $\gamma$ for different movement rates.
Figures (a--c) have low group tolerance (LGT).
`Analytic (LGT)' plot is calculated using equation \eqref{eq_analytic}, which represents the low migration limit.
Figures (e--f) have high group tolerance (HGT).
}
\end{minipage}

\section{Discussion}
In this paper we have proposed an evolutionary framework where the population is undated through individual birth, death and migration such that individuals migrate between a network of sites. 
This framework can be used to understand the effect of migration on an evolutionary process for different network topologies.
In other frameworks used to study the effect of network topology on evolution, migration is coupled with birth and death to keep the population size constant as in evolutionary graph theory (EGT) \citep{lieberman2005}, or combined with birth \citep{pattni2021}, which is akin to dispersal in plants or the spread of infectious disease.
Here, migration is a separate process and not coupled with birth or death allowing migration behaviour based on ecological interactions to be considered.
This is illustrated by applying the framework to construct a birth-death-migration model where migration is density dependent based on the Markov migration mechanism in \cite{pattni2018}.
In this mechanism, individuals move in response to competition and the parameter used to control this is called group tolerance.
Individuals who have high group tolerance are less sensitive to competition so are indifferent to staying or migrating.
Individuals with low group tolerance are highly sensitive to competition so prefer beneficial groups and will stay if so, otherwise they migrate.

The framework allows for overlapping evolutionary and ecological timescales but in this paper we focus on the rare mutation limit.
This means the population evolves in adaptive sweeps \citep{gerrish1998} where a mutant either fixates or goes extinct prior to another mutant arising.
This allows the effect of network structure on evolution to be measured in terms of the fixation probability of a mutant.
Since this approach is also used in models based on the EGT framework, comparisons can therefore be made with a well-established set of literature.
When moving away from the rare mutation limit, an alternative is the fixation probability in the presence of clonal interference \citep{pattni2021}.

As an initial example of a model using the framework described in this paper, we opted to use a binary density-dependent migration regimen. 
In the low group tolerance case, where individuals choose to stay when they are alone, every site is always occupied as it is assumed that individuals die through competition only.
In the high group tolerance case, where individuals move even when they are alone, a state where all sites are occupied is approached as the competition rate decreases.
This in turn enabled us to obtain analytical results in the low migration limit and allows us to make comparisons with models with fixed population size where all sites are always occupied \citep{lieberman2005,yagoobi2021}.
Generally, density-dependent migration that changes gradually with group compositions can be considered as done in \cite{pattni2017,erovenko2019}.
In these models, evolutionary steps (where the composition of the population changes) happen at fixed time points to keep the population size fixed.

By altering the competition rate we can change the size of the population.
A high competition rate implies that individuals exist in solitude as any pair of individuals meeting would result in a death. 
For low group tolerance in this scenario, individuals therefore prefer being alone. 
In this setting, we achieved a scenario that is akin to EGT \citep{lieberman2005} where each site is occupied by a single individual. 
Note that this does not mean that the dynamics are identical to EGT, however, this enabled us to make comparisons to obtain further insights.
In EGT, it is apparent that mutation dynamics (the process of a mutant appearing in the population) are not explicitly taken into account because it will break the fixed population size assumption (in a population size of $n$, the population size has to temporarily be $n+1$ to account for the appearance of a mutant). 
The effect of this is that the fixation probability with mutation dynamics is weighted by the probability of an initial mutant competing for a space within the population.
When competition is high, space comes at a premium and individuals would essentially be competing for a single site.
With lower competition, individuals would be competing for a share of space within a site. 
In EGT, the initial mutant does not compete with the resident on the site it is placed in, and is therefore guaranteed a space in the population.
The fixation probability in EGT is therefore greater than or equal to that obtained in our model.

In EGT and its various extensions \citep{pattni2017,yagoobi2021}, a mutant is introduced into the population through a mutant appearance distribution specifying which resident will be replaced by a mutant. Two distributions commonly used are a uniform distribution or a temperature-weighted distribution \citep{allen2014}.
One way to apply this scheme in populations with variable size, where there are multiple initial states, is to fix the initial state and then replace one resident with a mutant.
For example, the initial state can be fixed to one where each site is occupied by one resident \citep{pattni2021}.
This approach is sufficient for specific purposes, for example, understanding how EGT dynamics can be derived from a model with eco-evolutionary dynamics \citep{pattni2021}.
However, for the rare mutation limit, the statistically correct method is to  consider the distribution of resident states in the rare mutation limit ($\mu\to0$).
This determines the  mutant appearance distribution in proportion to the number of resident individuals on a network site.
Since the mutant appearance distribution affects the average fixation probability \citep{allen2014,tkadlec2019}, it is advantageous to explicitly take into account the mutation dynamics so that results can be interpreted unambiguously.
In terms of simulated computations, mutation dynamics can reduce the number of simulations that are run to fixation if the fixation probability is lower, saving computation time.

For a low competition rate, individuals can coexist with one another and we can obtain a metapopulation model where there are multiple individuals per site.   
As a point of comparison for such a scenario we use  \cite{yagoobi2021}, which considers a metapopulation model with a fixed number of individuals per site.  
In \cite{yagoobi2021} it is observed that the circulation theorem \citep{lieberman2005} (circulation networks have the same fixation probability) holds for all migration rates, where as in our case this is observed for low and high migration rates.
This is because, in our case, the number of individuals on a site is given by a distribution that is determined by the dynamics of the resident population prior to a mutation occurring.
For intermediate migration rates, the distribution of individuals is largely dependent upon the dynamics between a local neighbourhood of sites, which would differ between networks due to their topology. 
This is not the case for low and high migration rates.
For low migration rates, the number of individuals in a site would depend upon the dynamics within a site, so if the dynamics are the same within a site the number of individuals in a site would be the same regardless of the network.
For high migration rates, the number of individuals in a site would depend upon dynamics across all sites as individuals are constantly interacting with one another, and therefore network structure has little effect.

In the star network, \cite{yagoobi2021} observes that the star network amplifies selection, which we observe as well for low migration rates.  
A uniform mutant appearance distribution is used in \cite{yagoobi2021} so mutants are likely to appear on leaf sites regardless of the migration rate.
For the birth-death-migration model we define, the distribution of individuals changes such that the number of residents present in the centre site increases.
Mutants are then more likely to appear in the centre, which suppresses selection.
\cite{yagoobi2021} also considered different subpopulation sizes in a two-patch metapopulation, which they showed is a suppressor.
We can implement sites with different population sizes by, for example, using site-specific competition rates or using networks with sink and source sites. 
The star network is an example of the latter; the centre site is a sink and leaf sites are sources, which means that more individuals are migrating from a leaf site to the centre site than vice versa.
As the migration rate increases, the number of individuals in the centre site increases, which suppresses the fixation probability.

In summary, we have presented a network structured population evolution framework where birth, death and migration are uncoupled.
We study the effect of network structure in the rare mutation limit and have shown how the mutant appearance distribution affects the success of an invading mutant.
Future work will move away the rare mutation limit so that overlapping evolutionary and ecological timescales can be considered in the context of network structure.

\section*{Acknowledgements}
KP and KS acknowledge funding from EPSRC project grant EP/T031727/1.
This project has received funding from the European Union’s Horizon 2020 research and innovation programme under grant agreement No 955708.

\appendix
\section{Simulation Details}
The evolutionary process is simulated using the Gillespie algorithm \cite{gillespie1976,gillespie1977}. 
Recall that the infinitesimal generator (equation \eqref{equ:Simple_Movement}) describing the evolutionary process is as follows
\begin{equation}
\begin{split}
\mathcal{L}\phi(\mathcal{S}) = 
    &\sum_{i\in \mathcal{S}} 
        [1 - \mu(i)]
        b(i,\mathcal{S})
        [\phi(\mathcal{S} \cup \{i\}) - \phi(\mathcal{S})] \\
    & + \sum_{i \in \mathcal{S}} 
        \mu(i)b(i,\mathcal{S}) 
        \int_{\mathbb{R}^{l}}
        [\phi(\mathcal{S}\cup \{(u,X_i)\}) - \phi(\mathcal{S})]M(U_i,u)du\\
    & + \sum_{i\in \mathcal{S}} d(i,\mathcal{S})[\phi(\mathcal{S}\backslash \{i\}) - \phi(\mathcal{S})]\\
    & +\sum_{i \in \mathcal{S}}\sum_{x\in \mathcal{X}} m(i,x,\mathcal{S})[\phi(\mathcal{S} \cup \{(U_i,x)\}\backslash\{i\} - \phi(\mathcal{S})].
    \nonumber
\end{split}    
\end{equation}
For this process, let $T(k)$ and $S(k)$ respectively be the time and state after $k$ events.
The simulation follows the following steps:
\begin{enumerate}
    \item The time, $T(k+1)$, when the next event happens is given by
    \begin{align}
        T(k+1)=T(k)- \frac{\ln(\text{Unif}(0,1))}{\lambda_k}
        \nonumber
    \end{align}
    where $\text{Unif}(0,1)$ is uniformly distributed random number in the range $(0,1)$ and 
    \begin{align}
        \lambda_k = \sum_{i\in S(k)}\sum_{x\in \mathcal{X}} b(i,S(k)) + d(i,S(k)) + m(i,x,S(k)).
        \nonumber
    \end{align}
    \item The next state, $S(k+1)$, is determined by:
    \begin{itemize}
        \item Birth without mutation: The probability that $I_i$ gives birth to an offspring with the same type is
            \begin{align}
                [1-\mu(i)] \frac{b(i,S(k))}{\lambda_k}
                \nonumber
            \end{align}
        then $S(k+1) = S(k) \cup \{(U_i,X_i)\}$.
        \item Birth with mutation: The probability that $I_i$ gives birth to an offspring with type $w$ is
            \begin{align}
                \mu(i)M(U_i,w)\frac{b(i,S(k))}{\lambda_k}.
                \nonumber
            \end{align}     
        then $S(k+1) = S(k) \cup \{(w,X_i)\}$. 
        \item Death: The probability that $I_i$ dies is
        \begin{align*}
            \frac{d(i,S(k))}{\lambda_k}
        \end{align*}
        then $S(k+1) = S(k) \setminus \{i\}$.
        \item Movement: The probability that $I_i$ moves to site $n$ is
        \begin{align}
            \frac{m(i,n,S(k))}{\lambda_k}
            \nonumber
        \end{align}
        then $S(k+1) = S(k) \cup\{(U_i,n)\} \setminus \{i\}$. 
    \end{itemize}
    \item Repeat steps 1 and 2 as required.
\end{enumerate}
For the birth-death-migration model to simulate the hitting probability (equation \eqref{eq_hitting_prob}), 
\begin{align*}
    \mathcal{L}h_\mathcal{M}(\mathcal{S}) = 0
\end{align*}
with boundary conditions $h_\mathcal{M}(\mathcal{S}) = 1$ for $\mathcal{S} \in \mathcal{M}$ and $h_\mathcal{M}(\mathcal{S}) = 0$ for $\mathcal{S} \in \mathcal{R}$, we first need to determine the initial state in the rare mutation limit. 
To do this, we set $T(0)=0$, $\mu(i)=1^{-4}\ \forall i$, $M(R,M)=1$ and choose $S(0)\in\mathcal{R}$ such that $S(0)$ is at the carrying capacity in the deterministic system.
This is an added step taken to ensure that the stochastic system is fluctuating around its carrying capacity prior to a mutant arising.
The deterministic system is obtained by assuming that, rather than there being a discrete number of individuals, the number of individuals is continuous. 
Let $e_1(t),\ldots,e_N(t)$ be the number of residents at time $t$ in each site. 
For low group tolerance (equation \eqref{eq_lgt}), we  want the solution to the system (dropping $t$ for brevity)  
\begin{align}
    \frac{\text{d}e_x}{\text{d}t}=
    \beta_R e_x 
    - \gamma e_x (e_x-1)
    + \sum_y \lambda (e_y-1) W_{y,x} - \lambda(e_x - 1)W_{x,y}
    = 0,
\end{align}
where the first term accounts for birth, second term death, third term is immigration and fourth term is emigration.
Note that migration takes place only if the number of individuals in a site is $>1$.
Similarly, for high group tolerance (equation \eqref{eq_hgt}) we want the solution to 
\begin{align}
    \frac{\text{d}e_x}{\text{d}t}=
    \beta_R e_x 
    - \gamma e_x (e_x-1)
    + \sum_y \lambda e_y W_{y,x} - \lambda e_x W_{x,y}
    = 0,
\end{align}
the terms are as in the low group tolerance case but migration in this case happens when the individuals in a site is $>0$.
After obtaining $e_1(t),\ldots,e_N(t)$, we round up to the nearest integer and set $S(0)$ to this.
We then repeat steps 1 and 2 as outlined above until a mutant appears. 
Once a mutant appears, we use this as the initial state to calculate the hitting probability.
To continue the simulation, we set $\mu(i)=0\ \forall i$ and continue the simulation until we hit a state in $\mathcal{R}$ or $\mathcal{M}$. 
This is one run of the simulation which we repeat to generate multiple simulations.
Note that initialising the population in this way takes into account the mutant appearance distribution ($p_{x,\mathcal{S}}$) since a mutant is more likely to appear in a site with more individuals. 
The average fixation probability of a mutant is then given by $N_\text{mut}/N_\text{sim}$ where $N_\text{sim}$ and $N_\text{mut}$ are the total number of simulations and the number of simulations that hit $\mathcal{M}$ respectively.

\newpage
\bibliographystyle{agsm}
\bibliography{Project1}

\end{document}